\documentclass[11pt]{JHEP3}

\preprint{NSF-KITP-03-35 \\ SISSA 46/2003/EP \\ hep-ph/0305322}

\keywords{Neutrino Physics, Baryogenesis}

\usepackage{amsmath,amssymb,bbold,epsfig}

\newcommand{\beq}{\begin{equation}}

\title{Probing the seesaw mechanism with neutrino data and leptogenesis}

\author{Evgeny Kh. Akhmedov 
\thanks{On leave from the National
Research Centre Kurchatov Institute, Moscow, Russia.}\\
The Abdus Salam International  Center for Theoretical Physics (ICTP)\\ 
Strada Costiera 11, I-34100 Trieste, Italy \\
E-mail: \email{akhmedov@ictp.trieste.it}}

\author{Michele Frigerio\\
INFN, Section of Trieste and International  School
for Advanced Studies (SISSA)\\
Via Beirut 4, I-34013 Trieste, Italy\\
E-mail: \email{frigerio@he.sissa.it}}

\author{Alexei Yu.\ Smirnov\\
The Abdus Salam International  Center for Theoretical Physics (ICTP)\\ 
Strada Costiera 11, I-34100 Trieste, Italy, and\\
Institute for  Nuclear Research,  Russian Academy of Sciences, 
Moscow, Russia\\
E-mail: \email{smirnov@ictp.trieste.it}}

\abstract{
In the framework of the seesaw mechanism with three heavy
right-handed Majorana neutrinos and no Higgs triplets
we carry out a systematic study of the structure of the right-handed
neutrino sector. Using the current low-energy neutrino data as an input
and assuming 
hierarchical Dirac-type neutrino masses $m_{Di}$,
we calculate the masses $M_i$ and the mixing of the
heavy neutrinos. We confront the inferred properties of these
neutrinos  with the constraints coming from the requirement of a
successful baryogenesis via leptogenesis.
In the generic case the masses of
the right-handed neutrinos are highly hierarchical:
$M_i \propto m_{Di}^2$; the 
lightest mass is $M_1 \approx 10^3 - 10^6$ GeV and the generated
baryon-to-photon ratio $\eta_B\lesssim 10^{-14}$ is much smaller than 
the observed value.
We find the special cases which correspond to the level crossing points, 
with maximal mixing between two quasi-degenerate right-handed neutrinos. 
Two level crossing conditions are obtained: ${m}_{ee}\approx 0$
(1-2 crossing) and $d_{12}\approx 0$ (2-3 crossing), 
where ${m}_{ee}$ and $d_{12}$ are respectively the $11$-entry 
and the $12$-subdeterminant of the light neutrino mass matrix in the 
basis where the neutrino Yukawa couplings are diagonal.
We show that sufficient lepton asymmetry can be produced only
in the 1-2 crossing where $M_1 \approx M_2 \approx 10^{8}$ GeV, 
$M_3 \approx 10^{14}$ GeV and $(M_2 -  M_1)/ M_2 \lesssim 10^{-5}$.
}

\begin{document}

\section{Introduction}

The seesaw mechanism of neutrino mass generation \cite{yana} 
provides a very simple and appealing explanation of the smallness of the 
neutrino mass. 
The low-energy neutrino mass matrix $m$ is given in terms of the 
Majorana mass matrix of the right-handed (RH) neutrinos, $M_R$, and the Dirac 
mass matrix, $m_D$, as 
\beq
m =- m_D M_R^{-1} m_D^T ~.
\label{seesaw}
\end{equation}
While the elements of $m_D$ are expected to be at or below the EW scale, 
the characteristic mass scale of RH neutrinos is naturally the GUT or 
parity breaking scale.
A very important feature of the seesaw mechanism, which makes it
even more attractive, is that it has a simple and elegant built-in
mechanism of production of the baryon asymmetry of the Universe --
baryogenesis through leptogenesis \cite{FY}. 

The main prediction of the seesaw mechanism is  the existence of RH Majorana 
neutrinos $N_i$. Being extremely heavy, these neutrinos are not 
accessible to direct experimental studies, though several indirect ways 
of probing the properties of the RH sector are known: 
\begin{itemize}
\item[-] studies of leptogenesis \cite{JPR};
\item[-] searches for signatures of Grand Unification;
\item[-] renormalization group running effects induced by RH neutrinos
({\it e.g.}, on the $b-\tau$ unification \cite{btau}). 
\end{itemize}
What can be learned about 
the heavy RH neutrino sector, using 
the currently available low-energy neutrino data as an input? The matrix $m$ 
in (\ref{seesaw}) can to a large extent be reconstructed from the available 
low-energy data; then Eq. (\ref{seesaw}) can be used to study $M_R$.

Obviously, for such an analysis one would also need to know the Dirac 
neutrino mass matrix $m_D$. Unfortunately, at present no information on $m_D$ 
is available, since there are almost no ways of studying it experimentally 
(though, in the context of certain SUSY models, 
some information on $m_D$ can be 
obtained from the studies of the rare decays like $\mu \rightarrow e \gamma$, 
$\tau \rightarrow e \gamma$ \cite{borzu}
\footnote{Using constraints coming from these lepton flavor violating decays 
together with neutrino data, one can, in principle, reconstruct the mass
matrices relevant for the seesaw. 
The consequences for leptogenesis are discussed, e.g.,
in \cite{davidR}.}). 
One is, therefore, forced to resort 
to some theoretical arguments. Such arguments are in general provided by the 
GUTs, in which $m_D$ is typically related at the unification scale to the mass 
matrices of up-type or down-type quarks or of charged leptons. Since all the 
quark and charged lepton masses are highly hierarchical, we will assume this 
to be true also for $m_D$. Moreover, in models in which $m_D$ is related to 
the up-type quark mass matrices and the mass matrix of charged leptons $m_l$, 
to the down-type quark matrix, one can expect that the left-handed rotations 
diagonalizing $m_l$ and $m_D$ are nearly the same. Their mismatch 
(described by the matrix $U_L$) is expected 
not to exceed the Cabibbo mixing in the quark sector ($U_L\sim U_{CKM}$). 
The large leptonic 
mixing observed in the low-energy  sector is then a consequence of the ``seesaw
enhancement of lepton mixing'' \cite{enha,tani} 
\footnote{There also exists a different class of seesaw models in which 
the large low-energy leptonic mixing is due to the large left-handed 
Dirac-type mixing \cite{baba}.}. 
Such an enhancement requires a strong (quadratic) mass hierarchy of
RH neutrinos and/or off-diagonal structure of $M_R$.

In this framework, studies of the structure of RH sector and leptogenesis
have been performed in a number of publications \cite{berger,fatra}.
It was realized that, due to a strong hierarchy of neutrino Dirac Yukawa
couplings (analogous to that of up-type quarks), the predicted baryon 
asymmetry is much smaller than the observed one, especially 
\cite{orloff,bra1} in the case of the LMA MSW solution of the solar 
neutrino problem. 
The asymmetry can be much larger if the hierarchy of Yukawa couplings is
similar to that of the down quarks or charged leptons and the largest
coupling is of order 1 \cite{falco}. In this case the hierarchy between the 
masses of RH neutrinos becomes weaker and, furthermore, large RH mixing can 
appear.

In this paper, we follow a bottom-up approach to reconstruct the RH
neutrino sector, under the assumption of hierarchical Dirac masses.
We perform a systematic study of all possible structures of the RH neutrino 
mass matrix consistent with the low energy neutrino data.
Although our general formalism is valid for an arbitrary left-handed
Dirac-type mixing, in most of our quantitative analysis we assume this 
mixing to be small; we comment on the opposite situation in the 
last section.
We study the dependence of the produced lepton asymmetry on the structure 
of $M_R$, calculating explicitly the RH mixing matrix $U_R$ and the relevant
CP-violating phases. We confirm that, in the generic case, too small an 
asymmetry is produced. We identify and study in detail the special cases
in which the observed baryon asymmetry can be generated.

The paper is organized as follows. In section \ref{frame} 
we formulate our framework. 
In section \ref{generic} we discuss the generic case when 
the three RH neutrinos have 
strongly hierarchical masses.
We give simple analytic expressions for the masses and the mixings 
of the RH neutrinos. In section \ref{cross} we describe the conditions 
under which the special cases are realized. They correspond to partial or 
complete degeneracy of the
RH neutrinos. In section \ref{special1} we analyze the special case which 
leads to a successful leptogenesis.
In sections \ref{special2} and \ref{special3} other special cases are 
described. Section \ref{con} contains discussion of our
results and conclusions.

\section{The framework \label{frame}}

\subsection{Low energy data and structure of light  
neutrino mass matrix \label{low}}
 
In the flavor basis ($\nu_e,\nu_\mu, \nu_\tau$) 
the Majorana mass matrix of light neutrinos, $m$, can be written 
in terms of the observables as 
\beq
m = U_{PMNS}^*m^{diag}U_{PMNS}^{\dag}~,
\label{matr}
\end{equation}
where $m^{diag} \equiv diag(m_1,~m_2,~m_3)$
and $U_{PMNS}$ \cite{ponte1} 
is the leptonic mixing matrix,  
for which we use the standard parameterization: 
\begin{equation}
U_{PMNS} =
\left( \begin{array}{ccc}
 c_{12}c_{13} & s_{12}c_{13} & s_{13} e^{-i\delta} \\ 
-c_{23}s_{12}-s_{23}c_{12}s_{13} e^{i\delta} &
 c_{23}c_{12}-s_{23}s_{12}s_{13} e^{i\delta} & s_{23}c_{13} \\ 
 s_{23}s_{12}-c_{23}c_{12}s_{13} e^{i\delta} & 
-s_{23}c_{12}-c_{23}s_{12}s_{13} e^{i\delta} & c_{23}c_{13}
\end{array} \right)\!\cdot K_0\,.
\label{U}
\end{equation}
Here $c_{ij} \equiv \cos \theta_{ij}$, $s_{ij} \equiv \sin \theta_{ij}$,
$\delta$ is the CP-violating Dirac phase and 
$$K_0=diag(e^{i\rho},1,e^{i\sigma})$$ 
contains the two CP-violating Majorana
phases. Except in a few cases, we will absorb $\rho$ and $\sigma$ in the light 
neutrino masses $m_i$, which are therefore allowed to be complex. 

{}From the solar, atmospheric, accelerator and reactor neutrino  
experiments we take  
the following  input (at $90\%$ C.L.) \cite{skatmo}: 
\beq
\begin{array}{l}
\Delta m^2_{sol}\equiv\Delta m^2_{12} =
\left( 7^{~ +10}_{~ -2} \right) \cdot 10^{-5}~{\rm eV}^2\;,~~~~
\tan^2 \theta_{12}=  0.42^{ + 0.2}_{ - 0.1} ;\\ 
\Delta m^2_{atm}\equiv\Delta m^2_{23}=
\left(2.5^{~ +1.4}_{~ -0.9} \right) \cdot 10^{-3} ~{\rm eV}^2\;,~~~~
\tan \theta_{23} = 1 ^{~+ 0.35}_{~- 0.25}; \\
\sin\theta_{13} \lesssim 0.2\;.  
\end{array}   
\label{data}
\end{equation}
The neutrinoless $2\beta$ decay experiments \cite{H-M} restrict 
the $ee$-element of the matrix $m$:
$$
|m_{ee}| < (0.35 \div 1.3) ~{\rm eV}~ \qquad(90\% ~{\rm C.L.})\,.
$$
The direct kinematic measurements give an upper bound on neutrino masses, 
relevant  in the case of degenerate mass  spectrum:
$m_{\nu_e} \approx |m_i| < 2.2$ eV \cite{mainz}.
However, the cosmological bound on the sum of light neutrino masses 
(at $95\%$ C.L.)~\cite{WMAP,han}, 
\beq
{\sum}_i |m_i|  < (0.7 \div 2.1) ~~{\rm eV}~,
\label{WMAP}
\end{equation}
leads to the strongest limit on individual masses:
$|m_i| < (0.23 \div 0.70)$ eV. 

Using this phenomenological information one can, to a large extent, 
reconstruct the mass matrix $m$, just by inserting the data (\ref{data}) - 
(\ref{WMAP}) into Eq. (\ref{matr}) \cite{small,M1,M2}. A significant freedom 
still exists due to the unknown absolute mass scale $|m_1|$ and CP-violating 
phases $\rho$, $\sigma$. The dependence of the structure of $m$ on the unknown 
$s_{13}$ and $\delta$ is weaker because of the smallness of $s_{13}$. 

In spite of the above-mentioned freedom, a generic feature of the mass 
matrix $m$ emerges: all its elements are of the same order (within a factor of 
10 or so of each other), except in some special cases. The reason for this 
is twofold: 
\begin{itemize}
\item the two large mixing angles $\theta_{12}$ and $\theta_{23}$;
\item a relatively weak hierarchy between the mass eigenvalues:
according to the LMA MSW solution of the solar neutrino problem, 
\beq
\frac{|m_2|}{|m_3|} \geq
\sqrt{\frac{\Delta m^2_{sol}}{\Delta m^2_{atm}}}  >  0.1 \div 0.15~.
\end{equation}
\end{itemize}
We will refer to the situation when all the matrix elements of $m$ are of 
the same order and there are no special cancellations as the generic case. 
The ``quasi-democratic'' structure of the mass matrix  
$m$ is the main starting point of our analysis; it has important 
implications for the seesaw mechanism and leptogenesis, as we will see in
section 3.

A strong hierarchy between certain matrix elements of $m$ can be realized
only for specific values of the absolute mass scale and CP-violating 
phases\footnote{Realistic neutrino mass textures 
with some exact zeros have been studied, {\it e.g.}, in \cite{FGM}.
Also the seesaw realizations of these zeros have been considered \cite{KKST}. 
}, and we will consider these special cases separately (sections 
\ref{special1}-\ref{special3}).

\subsection{Dirac mass matrix \label{sw}}

In the basis where the mass matrix of RH Majorana neutrinos is diagonal, the 
Dirac mass matrix can be written as  
\beq
m_D = U_L^{\dagger} m_D^{diag} U_R ~.
\label{dirac1}
\end{equation}
Here $U_L$ and $U_R$ are unitary matrices and  $m_D^{diag} \equiv 
diag(m_u, m_c, m_t)$, with the mass eigenvalues $m_{u,c,t}$ being real and
positive. We have denoted the eigenvalues of $m_D$ in analogy with up-type 
quark masses, but we do not require the exact coincidence of the quark and
leptonic masses. 
Our main assumption in this paper is that  there is a strong  hierarchy 
of the eigenvalues of $m_D$:   
\beq
m_u\ll m_c\ll m_t ~,
\label{HD}
\end{equation}
similar to the hierarchy of the quark or charged lepton masses. 
For numerical estimates, we will use the reference values
\beq
m_u=1 ~{\rm MeV}~,\qquad m_c=400 ~{\rm MeV}~,\qquad m_t=100 ~{\rm GeV}~,
\label{numeric}
\end{equation}
which approximately coincide with the up-type quark masses at the 
mass scale $\sim 10^{9}$ GeV \cite{qm}.

The matrix $U_L$ defined in Eq. (\ref{dirac1}) describes the mismatch between 
the left-handed rotations diagonalizing the charged lepton and neutrino 
Dirac mass matrices and, therefore, is the leptonic analogue of the quark 
CKM mixing matrix. It differs from the leptonic mixing matrix $U_{PMNS}$, 
defined in Eq. (\ref{matr}), which is probed in the low-energy neutrino 
experiments. The difference between $U_L$ and $U_{PMNS}$ is a consequence of 
the seesaw mechanism. By analogy with the CKM matrix where the mixing is 
small, one expects that the matrix $U_L$ is close to the unit matrix.

\subsection{Conditions for a successful leptogenesis \label{lepto}}

Let us consider the constraints on the seesaw parameters coming from the 
requirement of the successful thermal leptogenesis. We assume that  
a lepton asymmetry is generated by the CP-violating 
out-of-equilibrium decays of RH neutrinos \cite{FY}. 
The lepton asymmetry is then 
converted to a baryon asymmetry through the sphaleron processes \cite{KRS},
thus explaining the baryon asymmetry
of the Universe. We will use the recent experimental value of the 
baryon-to-photon ratio 
\cite{WMAP},
\beq
\eta_B = (6.5 \pm ^{0.4} _{0.3})\cdot 10^{-10} ~.
\label{etaB}
\end{equation}

The lepton number asymmetry, $\epsilon_i$, produced in the decay of a RH 
neutrino with the mass $|M_i|$ can be written as
\cite{luty}:
\beq
\epsilon_i = \dfrac{1}{8 \pi} \sum_{k\neq i} 
f\left(\dfrac{|M_k|^2}{|M_i|^2}\right) 
\dfrac{{\rm Im}[(h^\dag h)_{ik}^2]}{(h^\dag h)_{ii}}~. 
\label{epsi}
\end{equation}
Here $h$ is the matrix of neutrino Yukawa couplings in the basis where
$M_R$ is diagonal with real and positive eigenvalues. 
Using the relation $h \equiv m_D/v$ 
(where $v=174$ GeV is the electroweak VEV)  
and $m_D$ given in (\ref{dirac1}) we can write
\footnote{In SUSY models $v$ should be replaced with $v\sin\beta$. For
$\tan\beta\gtrsim 3$, this corresponds to a very small rescaling 
of Yukawa couplings.}
\beq
h^\dag h = \dfrac{1}{v^2} U_R^\dag (m_D^{diag})^2 U_R ~. 
\label{hh}
\end{equation}
We note in passing that in general we allow the non-zero elements $M_i$ 
of the diagonalized RH neutrino mass matrix $M_R^{diag}$ to be complex. 
However, the unitary matrix $U_R$ in Eq. (\ref{hh}) is defined in such a
way that it relates the basis where $m_D$ is diagonal to the basis 
where $M_R$ is diagonal with real and positive non-zero elements,  
i.e. the phases of $M_i$ should be included into the definition of $U_R$ 
(cf. Eqs. (\ref{rimi}) and (\ref{Phi}) below).
We will be always assuming the ordering $|M_1| \le |M_2| \le |M_3|$. 

In the standard electroweak model the function $f$ in Eq. (\ref{epsi}) is 
given by
\beq
f(x)=\sqrt{x}\left[
\dfrac{2-x}{1-x}-(1+x)\log\left(\dfrac{1+x}{x}\right)
\right] ~.
\label{f}
\end{equation}
This expression is valid for $||M_i|-|M_j|| \gg \Gamma_i+\Gamma_j$, where 
$\Gamma_i$ is the decay width of the $i$th RH neutrino, given at tree level by
$$
\Gamma_i = \dfrac{(h^\dag h)_{ii}}{8\pi} |M_i| ~.
$$
In the limit of the quasi-degenerate neutrinos ($x = |M_j/M_i|^2 
\rightarrow 1$), one formally obtains from (\ref{f})
\beq
f(x) \approx \frac{1}{1 - x} \approx \frac{|M_i|}{2(|M_i|-|M_j|)} 
\rightarrow \infty~.
\label{fdeg}
\end{equation}
However, in reality the enhancement of the asymmetry  
is limited by the decay widths $\Gamma_i$ and 
is maximized when $||M_i|-|M_j|| \sim \Gamma_i+\Gamma_j$ \cite{pila}.

The left-handed rotation $U_L$ does not enter the expression for the
lepton number asymmetry. Furthermore, $h^{\dagger} h$ is invariant under the 
transformation
\beq
U_R \rightarrow D U_R ~, 
\label{inv}
\end{equation}
where 
\beq
D = diag(e^{i\alpha},e^{i\beta}, e^{i\gamma})
\label{dmatr}
\end{equation}
and $\alpha$, $\beta$, $\gamma$ are arbitrary phases. Consequently, 
all the phases 
that can be removed from $U_R$ by the transformation (\ref{inv}) have no
impact on leptogenesis. 

The baryon-to-photon ratio can be written as \cite{BDP1}
$$
\eta_B \simeq 0.01 \sum_i \epsilon_i \cdot \kappa_i~,
$$
where the factors $\kappa_i$ describe the washout of the 
produced lepton asymmetry 
$\epsilon_i$ due to various lepton number violating processes. 
In the domain of the parameter space which is of interest to us, 
they depend mainly on the effective mass parameters 
\beq
\tilde{m}_i\equiv\dfrac{v^2(h^\dag h)_{ii}}{|M_i|} =
\dfrac{[U_R^\dag (m_D^{diag})^2 U_R]_{ii}}{|M_i|}~.
\label{tildem}
\end{equation}
For $10^{-2}~{\rm eV} <\tilde{m}_1<10^3~{\rm eV}$, the washout factor 
$\kappa_1$ can be well approximated by \cite{KT} 
\beq
\kappa_1 (\tilde{m}_1)\simeq 0.3 \left(\dfrac{10^{-3}~{\rm eV}}{\tilde{m}_1}
\right)
\left(\log\dfrac{\tilde{m}_1}{10^{-3}~{\rm eV}}\right)^{-0.6}~.
\label{k1}
\end{equation}
When $|M_1| \ll |M_{2,3}|$, only the decays of the lightest RH neutrino 
$N_1$ are relevant for producing the baryon asymmetry $\eta_B$, since the 
lepton asymmetry generated in the decays of the heavier RH neutrinos 
is washed out by the $L$-violating processes involving $N_1$'s, which are 
very abundant at high temperatures $T\sim |M_{2,3}|$. At the same time,
at $T\sim |M_1|$ the heavier neutrinos $N_2$ and $N_3$ have already
decayed and so cannot wash out the asymmetry produced in the decays of
$N_1$.

In Refs. \cite{DIM1,BDP1}, under the assumption $|M_1| \ll |M_{2,3}|$,  
the following absolute lower bound on the mass of the lightest RH 
neutrino was found from the condition of the successful leptogenesis: 
\beq
|M_1| \gtrsim 4\cdot 10^8 ~{\rm GeV}~.
\label{lowM1}
\end{equation}
The bound (\ref{lowM1}) corresponds to $\tilde m_1 \rightarrow 0$ and maximal 
$\epsilon_1$; for other values of $\tilde{m}_1$ and $\epsilon_1$ it is even 
stronger \cite{BDP1,BDP2}. Inequality (\ref{lowM1}) has been derived 
before the latest WMAP data became available \cite{WMAP}. These data (see 
Eq. (\ref{etaB})) strengthen the bound by a factor $\sim 1.5$: $|M_1|\gtrsim 
6\cdot 10^8$ GeV.

\subsection{Mass matrix of RH neutrinos}

Using the representation (\ref{dirac1}) for $m_D$, the matrix of
light neutrinos can be written as 
$$
m = - U_L^{\dagger} m_D^{diag} U_R (M_R^{diag})^{-1} U_R^T m_D^{diag} U_L^* ~. 
$$
Then, in the basis where 
\beq
m_D  = U_L^{\dagger} m_D^{diag}~,
\label{basis}
\end{equation}
the inverse mass matrix of the RH neutrinos equals   
\beq
W \equiv M_R^{-1} = U_R (M_R^{diag})^{-1} U_R^T, 
\label{W}
\end{equation}
and, correspondingly, the matrix $M_R$ itself is given by
\beq
M_R = U_R^* M_R^{diag} U_R^\dag ~.
\label{mr}
\end{equation}
From the seesaw formula one obtains, in the basis (\ref{basis}), 
\beq
W = - m_D^{-1} m (m_D^{-1})^T = 
- (m_D^{diag})^{-1} \hat{m} 
(m_D^{diag})^{-1} ~,
\label{mr-1}
\end{equation}
where 
\beq
\hat{m}\equiv U_L m U_L^T ~.
\label{tilm}
\end{equation}
When $U_L =\mathbb{1}$, that is, the Dirac left-handed rotation is  
absent, one has $\hat{m}= m$. 
When $U_L$ slightly deviates from $\mathbb{1}$ ({\it e.g.}, $U_L\approx 
U_{CKM}$), the difference between $\hat{m}$ and $m$ is within the 
present experimental uncertainty, apart from some particular cases. 
In most of our analysis we shall be assuming $U_L=\mathbb{1}$ 
and neglect the difference between $\hat{m}$ and $m$. All the analytic 
expressions that we derive for $U_L =\mathbb{1}$  are also valid for an 
arbitrary $U_L$, if one substitutes the matrix elements of ${m}$ by the 
corresponding elements of $\hat{m}$.

In supersymmetric scenarios, $U_L$ can be probed in lepton flavor 
violating decays like $\mu\rightarrow e\gamma$ or $\tau\rightarrow 
\mu\gamma$. If $U_L=\mathbb{1}$, these decays are strongly suppressed, 
whereas for $U_L\approx U_{CKM}$ one finds the predicted branching ratios 
to be close to the experimental upper bounds, provided that the 
slepton masses are of the order of $(100\div 200)$ GeV and the neutrino 
Dirac masses take the values given in Eq. (\ref{numeric}).
Therefore, if future experiments find a signal close to the present
upper bounds, this will not require large left-handed rotations and so 
will not invalidate our results.

Using $m_D^{diag}\equiv diag(m_u, m_c, m_t)$ in Eq. (\ref{mr-1}) and taking 
$\hat{m}= m$, we obtain the following symmetric matrix $W$:
\beq
W= - 
\left(\begin{array}{ccc}
\vspace*{0.2cm}
\dfrac{m_{ee}}{m_u^2} & \dfrac{m_{e\mu}}{m_um_c} & \dfrac{m_{e\tau}}{m_um_t}\\
\vspace*{0.2cm}
\dots & \dfrac{m_{\mu\mu}}{m_c^2} & \dfrac{m_{\mu\tau}}{m_cm_t}\\
\vspace*{0.2cm}
\dots & \dots & \dfrac{m_{\tau\tau}}{m_t^2} 
\end{array}\right) 
~.
\label{IMR}
\end{equation}
In what follows we will find the eigenvalues of $W$ and the
mixing matrix $U_R$ that diagonalizes $W$ according to Eq. (\ref{W}).

\section{The generic case \label{generic}}

As discussed in section \ref{low}, in general the matrix elements 
$m_{\alpha\beta}$ 
are all of the same order of magnitude. We have defined this situation as 
the generic case. It follows then from (\ref{IMR}) that the elements
of $W$ are highly hierarchical, with $W_{11}$ being by far the largest
one. Introducing for illustration the small expansion parameter
\beq
\lambda\sim\dfrac{m_u}{m_c}\sim\dfrac{m_c}{m_t}\sim 10^{-2}~,
\label{lambda}
\end{equation}
we obtain 
$$
W \sim -\dfrac{m_{ee}}{m_u^2} 
\left(\begin{array}{ccc}
1 & \lambda  & \lambda^2 \\
\dots & \lambda^2  & \lambda^3\\
\dots & \dots & \lambda^4 
\end{array}\right)~,
$$
where in each element factors of order $1$ are understood.

The largest eigenvalue of $W$ is given, to a very good approximation, by
the dominant $(11)$-element:
\beq
M_1\approx \dfrac{1}{W_{11}} = - \dfrac{m_u^2}{m_{ee}} ~.
\label{M1}
\end{equation}
The second largest eigenvalue of $W$ can be obtained from the dominant
$(12)$-block of the matrix (\ref{IMR}), just by dividing its determinant
by $W_{11}$. The mass $M_2$ is then the inverse of this eigenvalue:
\beq
M_2 \approx \dfrac{W_{11}}{W_{11}W_{22}-W_{12}^2} =
\dfrac{m_c^2 m_{ee}}{m_{e\mu}^2-m_{ee} m_{\mu\mu}}
~.
\label{M2}
\end{equation}
The smallest eigenvalue of $W$ can be found from the condition 
\beq
(m_um_cm_t)^2 = - m_1m_2m_3M_1M_2M_3 ~
\label{det}
\end{equation}
which is obtained by taking the determinants of both sides of
Eq. (\ref{seesaw}). This yields
\beq
M_3 \approx \dfrac
{m_t^2(m_{e\mu}^2 - m_{ee} m_{\mu\mu})} {m_1m_2m_3}~.
\label{M3}
\end{equation}
Thus, in the generic case the RH neutrinos have a very strong mass hierarchy: 
$M_1 \propto m_u^2$,  $M_2 \propto m_c^2$,  $M_3 \propto m_t^2$, in agreement
with the ``seesaw enhancement'' condition \cite{enha}. 

The matrix $W$ is diagonalized, to a high accuracy, by 
\beq
U_R \approx \left(
\begin{array}{ccc}
\vspace*{0.2cm}
1 & -\left(\dfrac{m_{e\mu}}{m_{ee}}\right)^*\dfrac{m_u}{m_c} &
\left(\dfrac{d_{23}}{d_{12}}\right)^*\dfrac{m_u}{m_t} \\
\vspace*{0.2cm}
\left(\dfrac{m_{e\mu}}{m_{ee}}\right)\dfrac{m_u}{m_c} & 1 &
-\left(\dfrac{d_{13}}{d_{12}}\right)^*\dfrac{m_c}{m_t} \\
\vspace*{0.2cm}
\left(\dfrac{m_{e\tau}}{m_{ee}}\right)\dfrac{m_u}{m_t} &
\left(\dfrac{d_{13}}{d_{12}}\right)\dfrac{m_c}{m_t} & 1 
\end{array}\right) \!\cdot K ~,
\label{rimi}
\end{equation}
where
$$
d_{23}\equiv m_{e\mu} m_{\mu\tau}-m_{\mu\mu}m_{e\tau}~,~~~~~
d_{13}\equiv m_{ee}m_{\mu\tau}-m_{e\mu}m_{e\tau}~,~~~~~
d_{12}\equiv m_{ee}m_{\mu\mu}-m_{e\mu}^2
$$
and
\beq
K=diag(e^{-i\phi_1/2},e^{-i\phi_2/2},e^{-i\phi_3/2})~,~~~~~
\phi_i \equiv \arg{M_i} ~.
\label{Phi}
\end{equation}
As can be seen from Eq. (\ref{rimi}), the RH mixing is very small in the 
generic case. We therefore encounter an apparently paradoxical situation, 
when both the left-handed and RH mixing angles are small and yet one arrives 
at a strong mixing in the low-energy sector. This is an example of the  
so-called ``seesaw enhancement'' of the leptonic mixing 
\cite{enha,tani}. 
The reason for this enhancement can be readily understood. Indeed, small
mixing in $m_D$ and $W$ is related to the hierarchical structures of these 
matrices; however, in the seesaw formula (\ref{seesaw}) these hierarchies
act in the opposite directions and largely compensate each other, leading
to a ``quasi-democratic'' $m$ and thus to large mixing in the low-energy 
sector.

The masses of the heavy neutrinos (Eqs. (\ref{M1}),(\ref{M2}) and (\ref{M3})) 
can be rewritten as functions  of the low-energy observables 
using the expressions of $m_{\alpha\beta}$ 
in terms of the masses and mixing of light neutrinos \cite{small,M1,M2}. 
In the limit $\theta_{13}=0$ and $\theta_{23}=\pi/4$, we find
from Eq.~(\ref{matr}) 
\beq
\begin{array}{l}
M_1=-\dfrac{m_u^2}{m_1 c_{12}^2+ m_2 s_{12}^2} ~,\\ \\
M_2=-\dfrac{2m_c^2(m_1 c_{12}^2+ m_2 s_{12}^2)}
{m_3(m_1 c_{12}^2+m_2 s_{12}^2)+m_1m_2} ~,\\ \\
M_3=-\dfrac{m_t^2[m_3 (m_1c_{12}^2+
m_2 s_{12}^2)+m_1m_2]}{2m_1m_2m_3} ~.
\end{array}
\label{masses}
\end{equation}
The dependence of $|M_i|$ on $|m_1|$ is shown in Fig. \ref{fig}. 
In the case of the normal hierarchy ($|m_1|\ll 
|m_{2,3}|$), these equalities take a particularly simple form 
(found previously in \cite{bra1}): 
\beq
M_1\approx-\dfrac{m_u^2}{m_2 s_{12}^2}~,~~~~~~
M_2\approx- \dfrac{2 m_c^2}{m_3}~,~~~~~~
M_3\approx-\dfrac{m_t^2 s_{12}^2}{2m_1}~.
\label{NH}
\end{equation}
Notice that the lightest RH neutrino mass $|M_1|$ is related  
to the solar mass squared difference ($|m_2|^2\approx\Delta m^2_{sol}$),
$|M_2|$ to the atmospheric one ($|m_3|^2\approx\Delta m^2_{atm}$), and
$|M_3|$ is inversely proportional to $m_1$, for which 
we can use the upper bound $|m_1| < \sqrt{\Delta m^2_{sol}}$. 
It is illuminating to rewrite Eq. (\ref{NH}) in the ``standard'' seesaw  
form, expressing the light neutrino masses through the heavy neutrino
ones: 
\beq
m_1\approx -\dfrac{m_t^2 s_{12}^2}{2M_3}~,\quad\quad
m_2\approx -\dfrac{m_u^2}{s_{12}^2 M_1}~,\quad\quad
m_3\approx -\dfrac{2 m_c^2}{M_2}~.
\label{NH1}
\end{equation}
Comparing this with the naive seesaw expectations, we see that the expected 
correspondence between the masses of the light neutrinos and the Dirac 
masses $(m_1\propto m_u^2,~m_2\propto m_c^2~,~m_3\propto m_t^2)$ is 
completely broken; this is due to the large neutrino mixing angles (in 
particular, to the fact that the solution of the solar neutrino problem is 
the LMA MSW one).

Numerically, from Eq. (\ref{NH}) we find
\begin{eqnarray}
|M_1| & \simeq & \frac{m_u^2}{s_{12}^2 \sqrt{\Delta m^2_{sol}}} ~\simeq~    
4.4 \cdot 10^5 ~{\rm GeV} \left(\frac{m_u}{1 ~{\rm MeV}}\right)^2\,,
\label{NHa} \\
~|M_2| & \simeq & \frac{2 m_c^2}{\sqrt{\Delta m^2_{atm}}} ~\simeq~
6.4 \cdot 10^9 ~{\rm GeV} \left(\frac{m_c}{400 ~{\rm MeV}}\right)^2\,,
\label{NHb} \\
|M_3| & \simeq &  \frac{m_t^2 s_{12}^2}
{2 |m_1|} ~>~ 1.8
\cdot 10^{14} ~{\rm GeV} \left(\frac{m_t}{100 ~{\rm GeV}}\right)^2\,.
\label{NHc} 
\end{eqnarray}
These values of $|M_i|$ are illustrated by the leftmost regions (corresponding 
to $|m_1|\rightarrow 0$) of the plots in Fig. \ref{fig}. For the inverted 
mass hierarchy ($|m_3|\ll |m_1|\simeq |m_2|$), from Eq. (\ref{masses}) one 
finds
\begin{eqnarray}
|M_1| & \simeq & (2 - 5)\cdot 10^4 ~~{\rm GeV}\,
\left(\dfrac{m_u}{1~{\rm MeV}}\right)^2~, 
\label{IHa} \\
|M_2| & \simeq & (3 - 6)\cdot 10^9 ~~{\rm GeV}\,
\left(\dfrac{m_c}{400~{\rm MeV}}\right)^2~,
\label{IHb} \\
|M_3| & > & 10^{14} ~~{\rm GeV}\,\left(\dfrac{m_t}{100~{\rm GeV}}\right)^2~.
\label{IHc}
\end{eqnarray}
Similar estimates hold true also in the quasi-degenerate case
($|m_1| \simeq |m_2|\simeq |m_3|\simeq m_0$), except that 
the inequality sign in Eq. (\ref{IHc}) has to be replaced by the approximate 
equality one and the right-hand sides of Eqs. (\ref{IHa}) - (\ref{IHc}) 
have to be divided by $m_0/\sqrt{\Delta m^2_{atm}}\approx 20m_0/$eV.
In particular, for the lightest of the RH neutrinos 
we obtain
\beq
|M_1| \simeq (1 - 2.5) ~{\rm TeV} \,
\left(\dfrac{m_u}{1~{\rm MeV}}\right)^2 \,
\left(\dfrac{1~{\rm eV}}{m_0}\right)~.
\label{QD1}
\end{equation}
For the highest allowed by cosmological observations value, $m_0=0.7$ 
eV, Eq. (\ref{QD1}) gives $|M_1|\simeq (1.4 - 3.5)$ TeV. The values of 
$|M_i|$ in the quasi-degenerate case are shown on the right-hand side 
(corresponding to $|m_1|\approx m_0\sim 1$ eV) of panels a, b, d in Fig. 
\ref{fig}.

We now turn to the discussion of leptogenesis in the generic case under the   
consideration.  Since the RH neutrino masses are highly hierarchical, the main 
contribution to the lepton asymmetry comes from the decays of the lightest RH 
neutrino $N_1$.  
{}From Eq. (\ref{NHa}) we find that, for $m_u\lesssim 10$ MeV,   
its mass is at least one 
order of magnitude smaller than the absolute lower bound (\ref{lowM1}). 
The normal hierarchy case is the most favorable one: for the other 
cases $|m_{ee}|$ is larger, leading to even smaller values of $|M_1|$.

Let us compute the value of $\eta_B$ in the generic case. 
{}From Eqs. (\ref{tildem}), (\ref{M1}) and (\ref{rimi}) we get
\beq
\tilde{m}_1\approx\dfrac{|m_{ee}|^2+|m_{e\mu}|^2+|m_{e\tau}|^2}{|m_{ee}|}
=\dfrac{|m_2|^2s_{12}^2+|m_1|^2c_{12}^2}{|m_2 s_{12}^2+m_1 c_{12}^2|}
+{\cal O}(s_{13})
~.
\label{tildem1}
\end{equation}
In the case of the hierarchical spectra of light neutrinos, this gives
\beq
\begin{array}{llr}
\tilde{m}_1\approx |m_2|\approx\sqrt{\Delta m^2_{sol}}~,~~ & 
\kappa_1(\tilde{m}_1)\approx  0.02 ~~~ &
{\rm (NH)~,}\\
\tilde{m}_1\approx \dfrac{|m_2|}{|s_{12}^2+e^{-2i\rho}c_{12}^2|}
\approx\dfrac{\sqrt{\Delta m^2_{atm}}}{\cos2\theta_{12} \div 1}~,~~ &
\kappa_1(\tilde{m}_1)\approx  0.001 \div 0.003 ~~~ & {\rm (IH)~}, 
\label{mtilde}
\end{array}
\end{equation}
where $\kappa_1$ has been estimated using Eq.~(\ref{k1}).
For $|M_1|\ll |M_{2,3}|$, the lepton asymmetry $\epsilon_1$, given by 
Eq. (\ref{epsi}), can be written as
\beq
\epsilon_1 \approx -\dfrac{3}{16 \pi} \left[ 
\dfrac{|M_1|}{|M_2|} \dfrac{{\rm Im}(h^\dag h)_{12}^2}{(h^\dag h)_{11}}+
\dfrac{|M_1|}{|M_3|} \dfrac{{\rm Im}(h^\dag h)_{13}^2}{(h^\dag h)_{11}}
\right]~. 
\label{epsiGC}
\end{equation}
{}From Eqs. (\ref{hh}) and (\ref{rimi}) we get
$$
v^2(h^\dag h)_{11} \approx m_u^2 I_{11}~,~~~~~
v^2(h^\dag h)_{12} \approx m_u m_c I_{12}~,~~~~~
v^2(h^\dag h)_{13} \approx m_u m_t I_{13}~,
$$
where $I_{ij}=I_{ij}(m_{\alpha\beta})$ are order $1$ coefficients.
Using these relations and expressions (\ref{M1}), (\ref{M2}) 
and (\ref{M3}) for $M_i$ in Eq. (\ref{epsiGC}) we find 
$$
\epsilon_1 = \dfrac{3}{16 \pi} \dfrac{m_u^2}{v^2} \cdot I(m_{\alpha\beta})~,
~~~~I(m_{\alpha\beta})\sim 1~.
$$
Then the produced baryon-to-photon ratio is given, up to a factor of order
one, by
$$
\eta_B \simeq 0.01 \cdot \epsilon_1 \cdot \kappa_1(\tilde{m}_1) \simeq 
4\cdot 10^{-16} \cdot 
\left(\frac{m_u}{1 ~{\rm MeV}}\right)^2 
\left(\dfrac{\kappa_1(\tilde{m}_1)}{0.02}\right)~.
$$
To reproduce the observed value of $\eta_B$, one would need $m_u \sim 1$ GeV.
Thus, a successful leptogenesis requires $m_u \sim m_c$, 
which contradicts our assumption of a strong hierarchy between the eigenvalues 
of $m_D$ and goes contrary to the simple GUT expectations.   

Our conclusions concerning the mass spectrum of RH neutrinos and the
generated baryon asymmetry in the generic case are in accord with 
those reached in the previous studies \cite{fatra,orloff,bra1,falco}.

\section{Special cases and level crossing \label{cross}}

The results of the previous section were essentially based on two 
assumptions: (1) $m_{ee} \neq 0$ and is of the order of other elements of 
$m$, so that the evaluations (\ref{M1}) and (\ref{M2}) of $M_{1,2}$ 
are valid, and (2) $m_{ee} m_{\mu\mu} -   m_{e\mu}^2 \neq 0 $,
so that the evaluations (\ref{M2}) and (\ref{M3}) of $M_{2,3}$ hold.
Let us analyze the situations when one of these conditions or both of them
are not satisfied.

Special case I: 
$$m_{ee} \rightarrow 0$$ or, equivalently, $W_{11} \rightarrow 0$.
Formally, Eqs. (\ref{M1}) and (\ref{M2}) imply that $|M_1| \rightarrow 
\infty$ and $|M_2|\rightarrow 0$ when $m_{ee} \rightarrow 0$. At some 
point (the ``level crossing'' point) they will become equal to each other. 
The approximate formulas in Eqs. (\ref{M1}) and (\ref{M2}) do 
not work when $m_{ee}$ becomes very small. In exact  calculations  
one gets a significant decrease of the level splitting and, therefore, 
strong mass degeneracy. This behavior can be seen in
Fig. \ref{fig}, where we show the dependence of the RH neutrino masses 
and of $|m_{ee}|$ on the lightest mass $|m_1|$, 
for different values of the Majorana phases of the light neutrinos $\rho$ 
and $\sigma$. Small value of $m_{ee}$ appears as a result of a cancellation
of different contributions, which can be realized only for $\rho = \pi/2$
(Fig. \ref{fig}, panels b and d). At the crossing points the mixing 
between the levels becomes maximal 
\footnote{The level crossing of RH neutrinos has been previously discussed 
in \cite{orloff}.}.

Special case II: 
$$
d_{12}\equiv  m_{ee} m_{\mu\mu} -   m_{e\mu}^2 \rightarrow 0~,
$$ 
or, equivalently, $(W_{11}W_{22}-W_{12}^2) \rightarrow 0$. 
In this limit, according to Eqs. (\ref{M2}) and (\ref{M3}), $M_{2}$ increases 
and  $M_{3}$ decreases, so that a crossing occurs between the $N_2$ and 
$N_3$ levels. At the crossing point the mixing becomes maximal.
In Fig. \ref{fig} we show the dependence of $|d_{12}|$ on $|m_1|$.
The crossing points coincide with zeros of the $(12)$-subdeterminant.
As we will see in section \ref{special2}, $|d_{12}|$ is a non-monotonous 
function of  $|m_1|$, so that, depending on the phases $\rho$ and 
$\sigma$, there can be zero (Fig. \ref{fig}a), one (Fig. \ref{fig}b) or two 
(Fig. \ref{fig}d) crossings of this type. 
For $\rho=0$, $\sigma=\pi/2$ and quasi-degenerate spectrum of 
light neutrinos (right-hand part of Fig. \ref{fig}c), $|d_{12}|$ 
is much smaller than the squares of the light neutrino masses. This leads 
to a quasi-degeneracy of $N_2$ and $N_3$ without level crossing.

Special case III: 
$$ m_{ee}\rightarrow 0 {\rm~~~and~~~} d_{12}\rightarrow 0 ~.$$
This is equivalent to the requirement that the elements $m_{ee}$ and 
$m_{e\mu}$ be both very small. 
In this case all three RH neutrino masses are of the same order.
The $1-2$ and $2-3$ crossing regions merge.

In Fig. \ref{fig2} we show the dependence of $|M_i|$, $|m_{ee}|$ and
$|d_{12}|$ on $|m_1|$ for non-zero $s_{13}$ and different values of the
Dirac phase $\delta$. Comparing Fig. \ref{fig}b and  Fig. \ref{fig2}, which
correspond to the same Majorana phases, we find that the effect of $s_{13}$
for zero $\delta$ (Fig. \ref{fig2}a) is reduced to a small shift of the 
crossing points. A different choice of the phase $\delta$ has more 
substantial effect: it can remove all crossings (Fig. \ref{fig2}b), 
remove only the $2-3$ crossing (Fig. \ref{fig2}c) or change the relative 
positions of the crossing points leading to quasi-degeneracy of all 
three RH neutrinos (Fig. \ref{fig2}d).

As one can see in Figs. \ref{fig} and \ref{fig2}, the generic case with a 
strong hierarchy and small mass of the lightest RH neutrino is realized 
practically in the whole parameter space, excluding the regions of the 
crossings. In general, with the increase of the overall scale of the 
light neutrino mass $|m_1|$, the masses of the RH neutrinos decrease.
 
In the  following sections we will consider the special cases in
detail.


\FIGURE[t]{
\epsfig{file=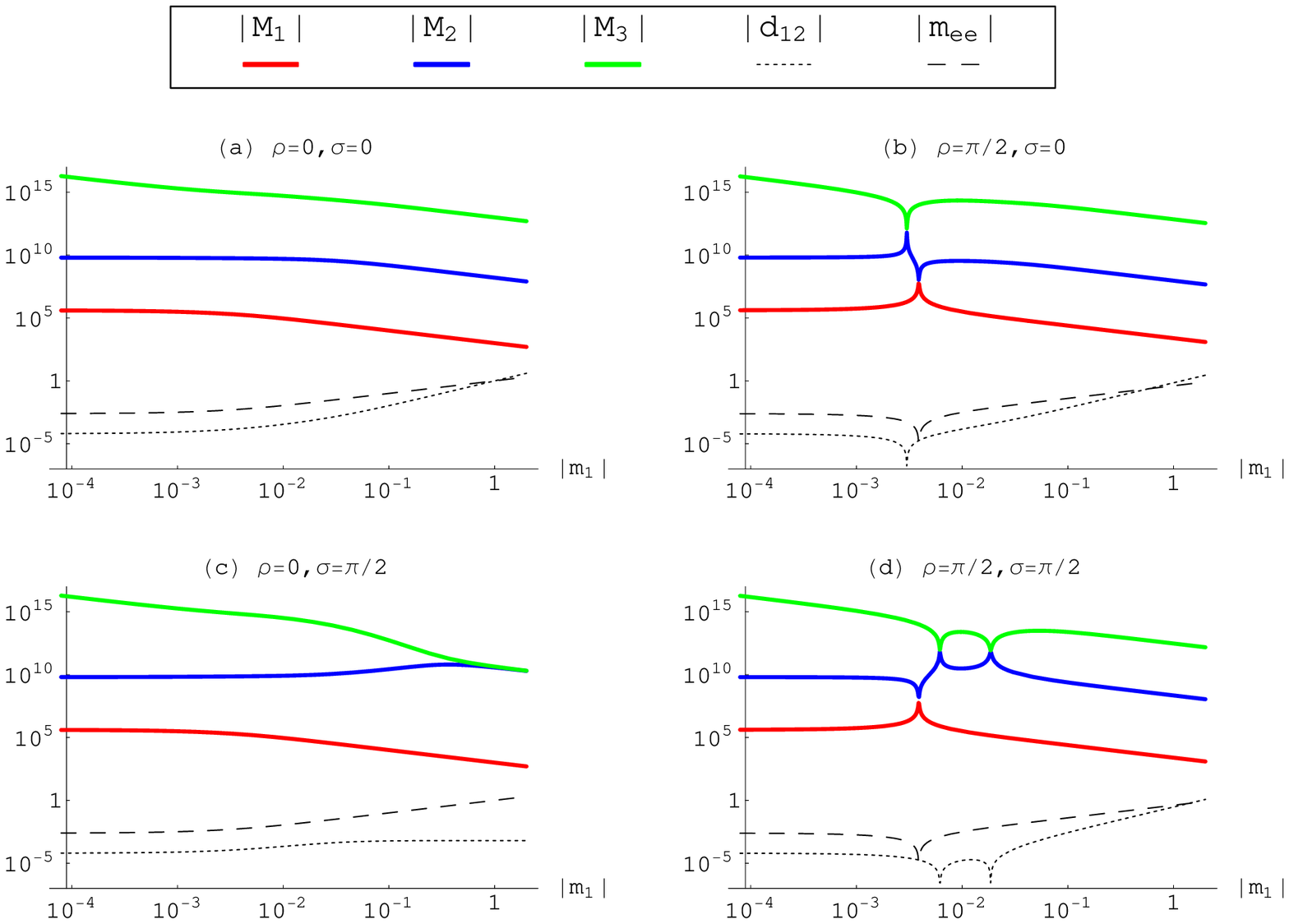,width=430pt,height=400pt}
\caption{
The masses of RH neutrinos $|M_i|$ in GeV as functions of the light neutrino
mass $|m_1|$ in eV (solid thick lines), 
for different values of the Majorana phases of light 
neutrinos, $\rho$ and $\sigma$. We have assumed normal mass ordering;
$s_{13}=0$; the best fit values of solar and atmospheric mixing angles and 
mass squared differences (Eq. (\ref{data})); the values of Dirac-type neutrino 
masses $m_{u,c,t}$ given in Eq. (\ref{numeric}). 
Also shown are $|d_{12}|\equiv |m_{ee} m_{\mu\mu}-m_{e\mu}^2|$ in eV$^2$ 
(dotted thin line) and $|m_{ee}|$ in eV (dashed thin line) as functions of 
$|m_1|$.
\label{fig}}}

\FIGURE[t]{
\epsfig{file=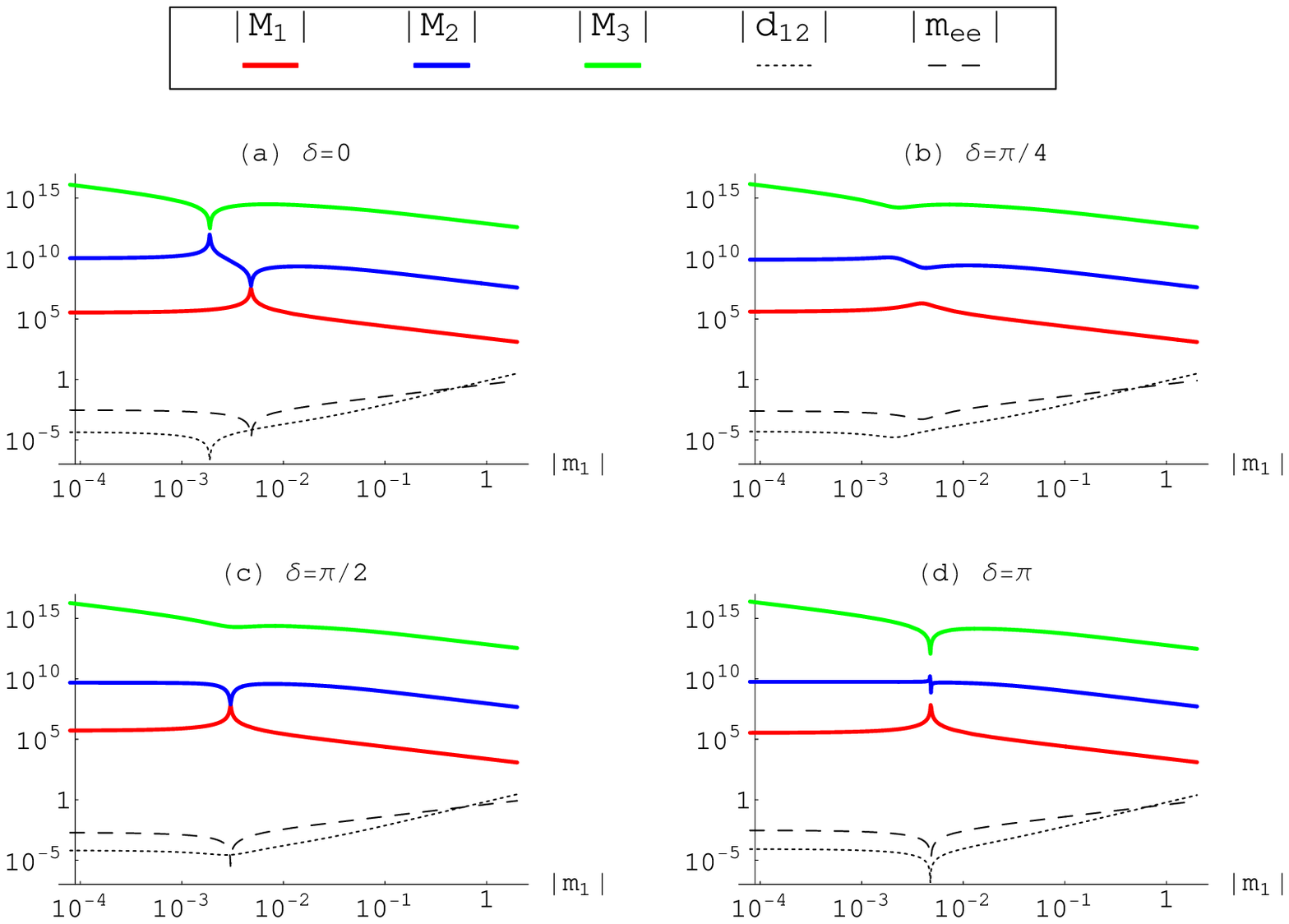,width=430pt,height=400pt}
\caption{
Same as in Fig. \ref{fig}, but for $\rho=\pi/2$, $\sigma=0$, $s_{13}=0.1$ 
and different values of the Dirac-type CP-violating phase $\delta$.
\label{fig2}}}

\section{Special case I: small $m_{ee}$ \label{special1}}  

Consider the case 
\beq
|m_{ee}| \ll \dfrac{m_u}{m_c} |m_{e\mu}| ~ 
\label{cond}
\end{equation}
which corresponds to  $|W_{11}| \ll |W_{12}|$ (see Eq. (\ref{IMR})). 
In this case the $(12)$-block of $W$ is 
dominated by the off-diagonal entries and, to a good approximation, the 
RH neutrino masses are 
\beq
M_1 \approx - M_2 \approx \dfrac{1}{W_{12}}\approx
-\dfrac{m_u m_c}{m_{e\mu}}~,~~~~~~
M_3\approx \dfrac{m_t^2 m_{e\mu}^2}{m_1m_2m_3} ~.
\label{S1}
\end{equation} 
Notice that $|M_1|$ is increased by a factor $\sim m_c/m_u$ with respect to
the generic case (Eq. (\ref{M1})).
Moreover, the RH (12)-mixing is nearly maximal while the other mixing 
angles are very small:  
\beq
U_R \approx \left(
\begin{array}{ccc}
\dfrac{1}{\sqrt{2}} & \dfrac{1}{\sqrt{2}} & 
\left(\dfrac{m_{\mu\mu}m_{e\tau}-m_{e\mu}m_{\mu\tau}}{m_{e\mu}^2}\right)^*
\dfrac{m_u}{m_t} 
 \\\\ 
-\dfrac{1}{\sqrt{2}}  & \dfrac{1}{\sqrt{2}} &
-\left(\dfrac{m_{e\tau}}{m_{e\mu}}\right)^*\dfrac{m_c}{m_t} \\\\
-\dfrac{m_{e\tau}}{\sqrt{2}m_{e\mu}}\dfrac{m_c}{m_t}  &
\dfrac{m_{e\tau}}{\sqrt{2}m_{e\mu}}\dfrac{m_c}{m_t} & 1 \\
\end{array}\right)\!\cdot K ~.
\label{rimiS1}
\end{equation}
The matrix of phases $K$ is given in Eq. (\ref{Phi}). 
Thus, the RH neutrinos $N_1$ and $N_2$ are 
quasi-degenerate, have nearly opposite CP parities and almost maximal mixing
($1-2$ level crossing).
The third RH neutrino $N_3$ is much heavier and weakly mixed with the 
first two.

Notice that, for $U_L=\mathbb{1}$,
$m_{ee}$ is the effective mass directly measurable 
in the neutrinoless $2\beta$ decay 
experiments. In our parameterization, it is given by
$$
m_{ee}=c_{13}^2(m_1 c_{12}^2+m_2 s_{12}^2)+s_{13}^2 e^{2i\delta} m_3 ~,
$$
so that  $m_{ee}\approx 0$ implies 
\beq
\tan^2\theta_{13}\approx -\dfrac{m_1 c_{12}^2+m_2 s_{12}^2}
{e^{2i\delta} m_3}~.
\label{zeroee}
\end{equation}
For $s_{13}=0$ the level crossing ($m_{ee}\rightarrow 0$) occurs for
\beq
|m_1|\approx \dfrac{\tan^2\theta_{12}\sqrt{\Delta m^2_{sol}}}
{\sqrt{1-\tan^4\theta_{12}}}
\approx (3-4)\cdot 10^{-3} {\rm eV} 
\label{m1mee}
\end{equation}
(see Fig. \ref{fig}, panels b and d). For substantial deviations of the 
$1-2$ mixing from the maximal one ($\tan^2\theta_{12}<1$), Eq. (\ref{m1mee}) 
can hold only in the case of the normal mass hierarchy. For the inverted
hierarchy or quasi-degenerate spectrum one has $|m_1|\gtrsim \sqrt{\Delta 
m^2_{atm}}\approx 0.05$ eV, so that Eq. (\ref{m1mee}) is not satisfied. 
Non-zero $s_{13}$ shifts the position of the level crossing. Taking into
account the present upper bound on $s_{13}$, we find that relation
(\ref{zeroee}) can be satisfied for $|m_1|\lesssim 0.02$ eV.
Moreover, the crossing takes place only for specific values of the
phase $\delta$ (see Fig. \ref{fig2}). 
If a stronger upper bound on $\theta_{13}$ is established, 
Eq. (\ref{zeroee}) will provide a more stringent upper bound on $|m_1|$ 
and also a lower bound on $|m_1|$.

Notice that inequality (\ref{cond}) implies
$$
|m_{ee}|<10^{-5}{\rm eV} \left(\dfrac{400m_u}{m_c}\right) ~,
$$
where we have taken $|m_{e\mu}|^2\lesssim\Delta m^2_{sol}$ 
(the normal hierarchy case).
If a positive signal is found in neutrinoless $2\beta$-decay experiments
with the near future sensitivity
($|m_{ee}|\gtrsim 0.01$ eV \cite{genius}), 
this special case will be excluded for $U_L=\mathbb{1}$.

Let us consider the effect of possible left-handed Dirac rotations 
assuming $U_L \sim U_{CKM}$. Taking for simplicity only a 1-2 rotation with
$\theta_L \sim \theta_c = 0.22$, we find from Eq. (\ref{tilm})
$$
\hat{m}_{ee} = \cos^2 \theta_L m_{ee} +
2\sin \theta_L \cos \theta_L m_{e\mu} +
\sin^2 \theta_L m_{\mu \mu}.
$$
Then the $1-2$ crossing condition, $\hat{m}_{ee} \rightarrow 0$, leads to 
the following restriction on the possible values of $m_{ee}$:
\beq
m_{ee} \approx - 2\tan \theta_L m_{e\mu} -  \tan^2 \theta_L m_{\mu \mu}\,,
\label{meecond}
\end{equation}
which can be considered as the level crossing condition in the flavor
basis. For the case of the normal mass hierarchy
$|m_{e\mu}| \sim 0.5 \sqrt{ \Delta m^2_{sol}}$ and
$|m_{\mu \mu}| \sim 0.5 \sqrt{ \Delta m^2_{atm}}$ and from
(\ref{meecond}) we find $|m_{ee}| \leq (3 - 4) \cdot 10^{-3}$ eV.

The level crossing condition (\ref{meecond}) can be satisfied also for 
the inverted mass hierarchy of light neutrinos as well as for the degenerate 
spectrum, if $\nu_1$ and $\nu_2$ have opposite CP parities. In the case of 
the inverted hierarchy one has $|m_{e\mu}| \sim 0.5 \sqrt{\Delta m^2_{atm}}> 
|m_{\mu \mu}|$ \cite{M2}, so that Eq. (\ref{meecond}) implies  
$|m_{ee}| \leq (1 - 2) \cdot 10^{-2}$ eV.
For the degenerate spectrum (taking into account the  cosmological
bound (\ref{WMAP})) one finds \cite{M1} $|m_{e\mu}| \sim |m_{\mu \mu}| 
\sim 0.2 - 0.4$ eV, and consequently $|m_{ee}| \leq 0.2$ eV.

In the limit $\theta_{13}=0$, we find from Eqs. (\ref{S1}) and (\ref{zeroee})
\begin{eqnarray}
|M_{1,2}| &\approx &  
\frac{2\sqrt{\cos2\theta_{12}}}{c_{23}\sin
2\theta_{12}} \frac{m_u m_c}{\sqrt{\Delta m_{sol}^2}}\approx 
9\cdot 10^{7} ~{\rm GeV} \left(\frac{m_u}{1 ~{\rm MeV}}\right)\!
\left(\frac{m_c}{400 ~{\rm MeV}}\right) \label{M12S1}\,, \\
|M_3| & \approx & \frac{c_{23}^2 m_t^2}{\sqrt{\Delta m_{atm}^2}}\approx 
10^{14} ~{\rm GeV} \left(\frac{m_t}{100 ~{\rm GeV}}\right)^2 \,
\label{M3S1}
\end{eqnarray} 
(see panels b and d of Fig. \ref{fig}).

Let us now consider the predictions for leptogenesis. 
Since $N_1$ and $N_2$ are quasi-degenerate and almost maximally
mixed, one expects nearly equal contributions to $\eta_B$ from their  
decays. Using Eqs. (\ref{S1}) and (\ref{rimiS1}), we find from (\ref{tildem})  
$$
\tilde{m}_1 \approx \tilde{m}_2 \approx
\dfrac{m_c}{m_u}\dfrac{|m_{e\mu}|^2+|m_{e\tau}|^2}{2|m_{e\mu}|}
=\dfrac{m_c}{m_u}\left(\dfrac{s_{12}c_{12}}{2c_{23}}|m_2 - m_1|
+{\cal O}(s_{13})\right)
~.
$$
Therefore, the maximal RH mixing in the $(12)$-sector leads to 
an increase of $\tilde{m}_1$ by a factor $\sim m_c/m_u$ with respect to
the generic case (Eq. (\ref{tildem1})) and, as a consequence, to a strong
enhancement of the washout effects. Taking into account Eq. (\ref{zeroee}), 
we obtain for $s_{13}\approx 0$
\beq
\tilde{m}_1\approx \dfrac{m_c}{m_u}\dfrac{\sin 2\theta_{12}
\sqrt{\Delta m^2_{sol}}}
{4c_{23}\sqrt{\cos 2\theta_{12}}}
\approx 1.5 ~{\rm eV}\left(\dfrac{m_c}{400m_u}\right)~,
\label{m1s1}
\end{equation}
and then, according to Eq. (\ref{k1}), 
\beq
\kappa_1(1.5~{\rm eV})
\approx  6\cdot 10^{-5}~.
\label{k1s1}
\end{equation}
The washout effects for lepton asymmetries produced in the decays of 
$N_1$ and $N_2$ are nearly the same: 
$\kappa_1(\tilde{m}_1)\approx \kappa_2(\tilde{m}_2)$.

Substituting the mixing parameters given by Eq. (\ref{rimiS1}) into 
Eq. (\ref{hh}), we find the relevant entries of $(h^\dag h)$:
\beq
\begin{array}{l}
(h^\dag h)_{12} \approx (h^\dag h)^*_{21} \approx 
-\dfrac{1}{2}\dfrac{m_c^2}{v^2}\left(1+\left|\dfrac{m_{e\tau}}{m_{e\mu}}
\right|^2\right) e^{i(\phi_1-\phi_2)/2} ~,\\
(h^\dag h)_{11} \approx (h^\dag h)_{22} \approx
\dfrac{1}{2}\dfrac{m_c^2}{v^2}
\left(1+\left|\dfrac{m_{e\tau}}{m_{e\mu}}\right|^2\right) ~,\\
(h^\dag h)_{13} \approx e^{i(\phi_1-\phi_2)/2}(h^\dag h)_{23} \approx
-\dfrac{1}{\sqrt{2}}\dfrac{m_c m_t}{v^2}
\left(\dfrac{m_{e\tau}}{m_{e\mu}}\right)^* e^{i(\phi_1-\phi_3)/2} ~. 
\end{array}
\label{hhS1}
\end{equation}
Then the contribution to $\epsilon_{1,2}$ coming from the diagrams with
the heaviest RH neutrino $N_3$ in the loop
(terms with $k=3$ in Eq. (\ref{epsi})) can be estimated as
follows:
$$
\epsilon_{1,2}^{(N_3)}\sim \pm\dfrac{3}{16\pi} \dfrac{m_u m_c}{v^2}
\sqrt{\dfrac{\Delta m^2_{atm}}{\Delta m^2_{sol}}}
\approx \pm 5\cdot 10^{-9} ~.
$$
These asymmetries are tiny compared to the values required for a successful 
leptogenesis and, moreover, have opposite signs for
$\epsilon_1$ and $\epsilon_2$. 
Therefore the dominant contribution should come from the diagrams with
$N_2$ ($N_1$) in the loop for the decay of $N_1$ ($N_2$). 
The corresponding asymmetries $\epsilon_{1,2}$ can be written as 
\beq
\epsilon_1\approx\epsilon_2\approx\dfrac{1}{16\pi}\dfrac{|M_1|}{|M_1|-|M_2|}
\dfrac{{\rm Im}[(h^\dag h)_{12}^2]}{(h^\dag h)_{11}}\approx
\dfrac{1}{32\pi}\dfrac{m_c^2}{v^2}
\left(1+\left|\dfrac{m_{e\tau}}{m_{e\mu}}\right|^2\right)\xi
~,
\label{epsigo}
\end{equation}
where
\beq
\xi=\dfrac{|M_1|}{|M_1|-|M_2|} \sin(\phi_1-\phi_2) ~.
\label{xi}
\end{equation}
The enhancement due to the quasi-degeneracy of $N_1$ and $N_2$ competes
with the suppression due to their almost opposite CP parities. 
Indeed, in the limit of exactly vanishing $W_{11}$ and $W_{22}$, one has 
$\sin(\phi_1-\phi_2)=\sin\pi=0$: in this case the complex phases can be 
removed by the transformation (\ref{inv}).
Taking into account terms of order $W_{11}$ and $W_{22}$, we
find
\beq
\xi\approx\dfrac{4k\tan\Delta}{(1+k)^2+(1-k)^2\tan^2\Delta}~,
\label{xi1}
\end{equation}
where
\beq
k\equiv\dfrac{|W_{22}|}{|W_{11}|}=\dfrac{m_u^2|m_{\mu\mu}|}{m_c^2 
|m_{ee}|}~,~~~~~
\Delta\equiv\dfrac 12 \arg\dfrac{W_{12}^2}{W_{11}W_{22}}=
\dfrac 12 \arg\dfrac{m_{e\mu}^2}{m_{ee}m_{\mu\mu}}~.
\label{k}
\end{equation}
Notice that the phase $\Delta$ is invariant under the transformation 
(\ref{inv}). For $|1-k|\ll 1/\tan\Delta$, Eq. (\ref{xi1}) gives  
$$
\xi\approx\tan\Delta ~,
$$ 
and for $\Delta\simeq \pi/2$ a significant enhancement of the asymmetries 
$\epsilon_{1,2}$ can be achieved. 
The enhancement factor depends on the degree of near-equality of 
$|W_{22}|$ and $|W_{11}|$.

For $k\rightarrow 1$, the level splitting can be written as
\beq
\dfrac{|M_2|-|M_1|}{|M_1|}\approx 2\dfrac{|W_{22}|}{|W_{12}|}\cos\Delta
= 2 \dfrac{|m_{\mu\mu}|}{|m_{e\mu}|} \frac{m_u}{m_c}\cos\Delta~,
\label{split}
\end{equation}
so that for $\Delta\approx \pi/2$ the splitting is substantially reduced.  
As we discussed in section \ref{lepto}, the enhancement due to the degeneracy
is restricted by the condition 
\beq
\dfrac{|M_2|-|M_1|}{|M_1|}\gtrsim \dfrac{\Gamma_1}{|M_1|}~,
\label{gam1}
\end{equation}
where in the case under the consideration
\beq
\dfrac{\Gamma_1}{|M_1|}\approx \dfrac{1}{8\pi}\dfrac{m_c^2}{2v^2} 
\left(1+\left|\dfrac{m_{e\tau}}{m_{e\mu}}\right|^2\right)~.
\label{gam2}
\end{equation}
Estimating $|m_{e\mu}|\approx |m_{e\tau}|\approx |m_{\mu\mu}|\sqrt{\Delta 
m^2_{sol}/\Delta m^2_{atm}}$, from Eqs. (\ref{split}), (\ref{gam1}) and 
(\ref{gam2}) we find the maximal possible enhancement:
\beq
\xi^{max}\approx \tan\Delta \approx \dfrac{1}{\cos\Delta}
\approx \dfrac{16\pi v^2 m_u}{m_c^3}
\sqrt{\dfrac{\Delta m^2_{atm}}{\Delta m^2_{sol}}}
\approx 1.4\cdot 10^5 ~. 
\label{tanDelta}
\end{equation}
In the numerical estimate we have taken the values in Eq. (\ref{numeric}). 
Using Eqs. (\ref{split}), (\ref{gam2}) and
(\ref{epsigo}), we can write the maximal asymmetry as
$$
\epsilon^{max}=\dfrac 12 \dfrac{\Gamma_1}{|M_1|} \xi^{max}
\approx  \dfrac{|W_{22}|}{|W_{12}|}\approx \dfrac{m_u}{m_c}
\sqrt{\dfrac{\Delta m^2_{atm}}{\Delta m^2_{sol}}}
~,
$$
which shows that $\epsilon^{max}\sim 10^{-2}$ is reachable in this scenario.

Combining Eqs. (\ref{k1}), (\ref{epsigo}) and (\ref{tanDelta}), we find 
$$
\eta_B \approx 0.01 \cdot 2 \epsilon_1 \kappa_{1}
\approx 1.9\cdot 10^{-8}
\left(\dfrac{400m_u}{m_c}\right)^2
\left[1+0.14\log\left(\dfrac{m_c}{400m_u}\right)\right]^{-0.6}
\left[\dfrac{\xi}{\xi^{max}(m_u,m_c)}\right]
~.
$$
Therefore the value (\ref{etaB}) of $\eta_B$ can be obtained for
$m_u/m_c \gtrsim 2\cdot 10^{-3}$. 
For $m_c=400m_u$, the observed baryon asymmetry
is reproduced for  
$\xi\approx\tan\Delta\approx 5\cdot 10^3$, which corresponds 
to the relative splitting (see Eq. (\ref{split}))
$$
\dfrac{|M_2|-|M_1|}{|M_1|}\approx 6\cdot 10^{-6}~.
$$

Thus, in spite of strong washout effects, a sufficiently large baryon 
asymmetry can be generated in this case, due to the enhancement related to 
the strong degeneracy of the RH neutrinos. For this to occur, not only 
the level crossing condition ($m_{ee}\rightarrow 0$) has to be satisfied, 
but also a special phase condition leading to $\Delta\approx\pi/2$ should be 
fulfilled. This value of $\Delta$ is consistent with the low energy neutrino 
data. We have checked the analytic results presented in this section by 
precise numerical calculations.

\section{Special case II: small $12$-subdeterminant of $m$ \label{special2}}

Let us consider the case in which the $(11)$-element of the matrix $W$ in Eq. 
(\ref{IMR}) is still the dominant one (as in the generic case), but the 
$(12)$-subdeterminant of $W$ is very small. Then 
$(M_R)_{33}$, which is proportional to this subdeterminant, is suppressed. 
The condition $(M_R)_{33}\ll(M_R)_{23}$ can be written as  
\beq
|d_{12}|\equiv|m_{ee}m_{\mu\mu}-m_{e\mu}^2|\ll \dfrac{m_c}{m_t}
|m_{e\tau}m_{e\mu}-m_{ee}m_{\mu\tau}| ~.
\label{small12}
\end{equation}
In this case $M_1$ is still given by Eq. (\ref{M1}), but $M_2$ cannot be 
found from the determinant of the $(12)$-block of $W$ as in Eq. (\ref{M2}). 
One has to consider, instead, the $(23)$-block of $M_R$, which is dominated by 
its off-diagonal entry. This yields  
\beq
M_2 \approx - M_3 \approx (M_R)_{23}= \dfrac{m_c m_t}{m_1m_2m_3} 
(m_{ee}m_{\mu\tau}-m_{e\tau}m_{e\mu}) ~.
\label{M23}
\end{equation}

The mixing matrix of RH neutrinos equals 
\beq
U_R \approx \left(
\begin{array}{ccc}
1 &
-\dfrac{1}{\sqrt{2}}\left(\dfrac{m_{e\mu}}{m_{ee}}\right)^*\dfrac{m_u}{m_c} &
-\dfrac{1}{\sqrt{2}}\left(\dfrac{m_{e\mu}}{m_{ee}}\right)^*\dfrac{m_u}{m_c} \\
\left(\dfrac{m_{e\mu}}{m_{ee}}\right)\dfrac{m_u}{m_c} & 
\dfrac{1}{\sqrt{2}} & \dfrac{1}{\sqrt{2}} \\
\left(\dfrac{m_{e\tau}}{m_{ee}}\right)\dfrac{m_u}{m_t} &
-\dfrac{1}{\sqrt{2}} & \dfrac{1}{\sqrt{2}} \\
\end{array}\right)\!\cdot K~,
\label{rimiS2}
\end{equation}
where $K$ is given in Eq. (\ref{Phi}). 
{}From Eq. (\ref{M23}) it follows that the phases of $M_2$ and $M_3$ differ 
by $\approx \pi$. 
Therefore in this special case the lightest RH neutrino 
is weakly mixed with $N_2$ and $N_3$, which are much heavier, 
quasi-degenerate, 
almost maximally mixed and have nearly opposite CP parities.

Let us consider condition (\ref{small12}). In terms of low-energy 
neutrino parameters, we obtain from Eq. (\ref{matr})
$$
d_{12}=
c_{23}^2 m_1 m_2 + s_{23}^2 m_3 
(c_{12}^2 m_1 + s_{12}^2 m_2)
+ {\cal O}(s_{13})~.
$$
Neglecting ${\cal O}(s_{13})$ corrections, we find 
that $d_{12}$ vanishes for
\beq
m_3=-\dfrac{\cot^2\theta_{23}m_1 m_2}
{c_{12}^2 m_1 + s_{12}^2 m_2}~.
\label{subdet}
\end{equation}
Since $\cot^2\theta_{23}\approx 1$, this relation cannot be satisfied for 
$|m_3|< |m_{1,2}|$. Therefore this special case is not realized for the 
inverted ordering of the light neutrino masses, unless $U_L$ deviates 
significantly from $\mathbb{1}$. For the normal mass ordering, condition
(\ref{subdet}) requires $|m_1|\gtrsim 2\cdot 10^{-3}$ eV
(see Fig. \ref{fig}, panels b and d). For the 
quasi-degenerate mass spectrum, Eq. (\ref{subdet})
can be satisfied if the CP parity of 
$\nu_3$ is opposite to the CP parities of 
$\nu_1$ and $\nu_2$, up to deviations of $\theta_{23}$ from $\pi/4$
(see the right-hand side of Fig. \ref{fig}c).
Notice that
$$
\begin{array}{rl}
m_{e\tau}m_{e\mu}-m_{ee}m_{\mu\tau} & =
s_{23}c_{23} [m_1 m_2 - m_3 (c_{12}^2 m_1 + s_{12}^2 m_2)]
+ {\cal O}(s_{13}) =\\
& = \cot\theta_{23} m_1 m_2 + {\cal O}(s_{13})~,
\end{array}
$$
where we have used the condition (\ref{subdet}) of zero $(12)$-subdeterminant.
Therefore Eq. (\ref{M23}) simplifies to
\beq
M_2\approx -M_3 \approx -\dfrac{m_c m_t}{m_3}~.
\label{massIa}
\end{equation}
Numerically, one finds (see Fig. \ref{fig}, panels b, c, d)
\beq
|M_2|\approx |M_3|\approx \frac{m_c m_t}
{(0.05 - 0.7){\rm eV}} 
\approx (0.6 - 8)\cdot 10^{11}~{\rm GeV} \left(\frac{m_c}{400 ~{\rm MeV}}
\right)\!\left(\frac{m_t}{100 ~{\rm GeV}}\right)~,
\label{massIb}
\end{equation}
while $|M_1|$ is still given by Eqs. (\ref{NHa}) and (\ref{QD1}) for the 
normal hierarchy and quasi-degenerate case, respectively. 

For this special case, the predictions for the lepton asymmetry 
are analogous to
those in the generic case. The production of the asymmetry is
dominated by the decays of the lightest RH neutrino. 
Due to the larger RH mixing, the asymmetry $\epsilon_1$ gets an enhancement 
factor $\sim (m_t/m_c)$ with respect to the generic case, but the leading 
terms in Im$(h^\dag h)_{12}^2$ and Im$(h^\dag h)_{13}^2$ cancel
because of the nearly opposite CP parities of $N_2$ and $N_3$. Indeed,
the sum of the two terms is proportional to $\sin(\phi_2-\phi_3)\approx 0$, 
where $\phi_i$ are defined in Eq. (\ref{Phi}). Thus, the produced lepton 
asymmetry is insufficient for a successful baryogenesis through leptogenesis. 
This is in agreement with the fact that in this special case the value of 
$|M_1|$ is still below the absolute lower bound (\ref{lowM1}).

\section{Special case III: small $m_{ee}$ and $m_{e\mu}$ \label{special3}}

Consider now the case when 
\beq
|m_{ee}|\ll\dfrac{m_u}{m_t}|m_{e\tau}|~,~~~~~
|m_{e\mu}|\ll\dfrac{m_c}{m_t}|m_{e\tau}|,~~\dfrac{m_u}{m_c}|m_{\mu\mu}|~,
\label{condi3}
\end{equation}
so that the $(13)$- and $(22)$-elements of $W\equiv M_R^{-1}$ are 
the dominant ones (see Eq. (\ref{IMR})).
In this case, two RH neutrinos form a quasi-degenerate pair with almost 
maximal mixing, opposite CP-parities and masses
\beq
\pm M_d\approx \pm W_{13}^{-1}\approx  \mp\dfrac{m_u m_t}{m_{e\tau}} ~.
\label{S33}
\end{equation}
The third neutrino has small mixing with the other two (of order 
$m_u/m_c$ or $m_c/m_t$) and a mass
\beq
M_s \approx W_{22}^{-1}\approx -\dfrac{m_c^2}{m_{\mu\mu}} ~.
\label{S3}
\end{equation}
Since $m_u m_t \sim m_c^2$, all the three masses are of the same order
\footnote{Seesaw mass matrices which correspond to the degeneracy 
of all three RH neutrinos have been recently considered in 
\cite{FJ,gla2}.}.

Let us consider conditions (\ref{condi3}).
We have shown in section \ref{special1} that, assuming $U_L
\approx\mathbb{1}$, $m_{ee}$ can be very small only in the case of the normal 
hierarchy, when Eq. (\ref{zeroee}) can be satisfied. 
At the same time, for certain values of $m_1$ and $s_{13}$ and of the 
phases, the value of $m_{e\mu}$ can also be very small. For this to occur, 
the low-energy parameters should satisfy (see Fig. \ref{fig2}d) 
$$
|m_1| \approx |m_2| \tan^2\theta_{12} \approx 0.0035 ~{\rm eV}~,~~~~~~
s_{13} \approx \left|\dfrac{m_2}{m_3}\right| \tan\theta_{12} \approx 0.11~,
$$
where we used $|m_2|\approx\sqrt{\Delta m^2_{sol}}$ and $|m_3|\approx
\sqrt{\Delta m^2_{atm}}$. The other matrix elements of $m$ are also 
approximately determined by the conditions $m_{ee}\approx 0$, 
$m_{e\mu}\approx 0$:
$$
|m_{e\tau}|\approx \sqrt{2}\tan\theta_{12}|m_2| \approx 0.008 ~{\rm 
eV}~,~~~~~~
|m_{\mu\mu}|\approx|m_{\mu\tau}|\approx|m_{\tau\tau}|\approx \dfrac{|m_3|}{2}
\approx 0.025 ~{\rm eV}~.
$$
For the masses of the RH neutrinos one then finds 
\begin {eqnarray}
|M_d| &\simeq &  1.2\times 10^{10} ~{\rm GeV} \left(\frac{m_u}{1 
~{\rm MeV}} \right)\left(\frac{m_t}{100 ~{\rm GeV}} \right)~, \\
|M_s| & \simeq & 6.4\cdot 10^{9} ~{\rm GeV}\left(\frac{m_c}{400 ~{\rm 
MeV}} \right)^2~.
\end{eqnarray}

{}According to Eqs. (\ref{S33}) and (\ref{S3}), 
the mass spectrum of RH neutrinos is characterized by the ratio  
\beq
r\equiv\dfrac{|M_d|}{|M_s|}=
\dfrac{m_u m_t}{m_c^2} \left|\dfrac{m_{\mu\mu}}{m_{e\tau}}\right| =
1.9 \left(\dfrac{m_u}{1 ~{\rm MeV}}\right) 
\left(\dfrac{m_t}{100 ~{\rm GeV}}\right) 
\left(\dfrac{400 ~{\rm MeV}}{m_c}\right)^2 
\left|\dfrac{m_{\mu\mu}}{3 m_{e\tau}}\right|~.
\label{r}
\end{equation}
We shall distinguish two subcases, depending on whether the
quasi-degenerate pair is lighter or heavier than the singlet state:
\begin{itemize}
\item[(a)]  $r<1 ~\Rightarrow~ |M_1| \approx |M_2| \lesssim |M_3|$ .
\item[(b)] $r>1 ~\Rightarrow~ |M_1|\lesssim |M_2|\approx |M_3|$ .
\end{itemize}
For certain values of the parameters, the splitting between the 
quasi-degenerate neutrinos can become larger than the 
difference between the masses of one of them and of the third neutrino. 
This case can be considered as the limit in which (a) and (b) merge. 
Notice that, in this limit, the structure of $U_R$ is very unstable, and 
all three RH mixing angles can be large.

Let us consider now the predictions for leptogenesis. To compute the produced
lepton asymmetry one has to take into account the interplay among all
three quasi-degenerate RH neutrinos.
The effects related to mass degeneracy and large RH mixing angles, discussed 
in section \ref{special1}, are present also here.
Notice that the maximal RH mixing is now related with the Dirac masses 
$m_u$ and $m_t$ rather than with $m_u$ and $m_c$, as in section \ref{special1}.
Let us discuss the two subcases defined above.

(a) $r<1$, light quasi-degenerate pair. 

Up to ${\cal O}(\lambda^2)$ terms, the RH mixing matrix is given by
\beq
U_R\approx\left(
\begin{array}{ccc}
1/\sqrt{2} & 1/\sqrt{2} & \delta_1^* \\
\dfrac{1}{\sqrt{2}}(\delta_2^* - \delta_1) & 
-\dfrac{1}{\sqrt{2}}(\delta_2^* + \delta_1) & 1 \\
-1/\sqrt{2} & 1/\sqrt{2} & \delta_2
\end{array}
\right) \!\cdot K' ~.
\label{ura}
\end{equation}
Here 
\beq
\delta_1 \equiv \dfrac{m_u}{m_c} ~\dfrac{m_{\mu\tau}}{m_{e\tau}}
~\dfrac{1}{r^2-1}~,~~~~~~
\delta_2 \equiv \dfrac{m_u}{m_c} \left(\dfrac{m_{\mu\tau}}{m_{e\tau}}\right)^*
\dfrac{r}{r^2-1}~,
\label{deltas}
\end{equation}
and $K'=diag(e^{-i\phi_1'/2},~e^{-i\phi_2'/2},~e^{-i\phi_3'/2})$, where 
\beq
\phi_1'-\phi_2' \approx \pi~,~~~~~~~\phi_3' - \phi_2' \approx 
2\arg\left(\dfrac{m_{e\tau}}{m_{\mu\tau}}\right)~.
\label{phases}
\end{equation}
Notice that in the parameterization (\ref{ura}) the phases $\phi_i'$ are
not the arguments of $M_i$. 

{}From Eqs. (\ref{S33}), (\ref{ura}) and (\ref{tildem}) we find 
the effective mass parameter
$$
\tilde{m}_1\approx\tilde{m}_2\approx\dfrac{m_t}{2m_u}|m_{e\tau}|
\approx 500 ~{\rm eV} ~,
$$
which leads, according to Eq. (\ref{k1}), to a very small washout factor
\beq
\kappa_{1,2}(500~{\rm eV})\approx 10^{-7}~.
\label{ti3}
\end{equation}
To survive such strong a washout, lepton asymmetries $\epsilon_i$ of order 
unity are required.

The contribution to $\epsilon_{1,2}$ of diagrams with $N_3$ in the loop 
can be estimated as
$$
\epsilon_{1,2}^{(N_3)}\sim \pm\dfrac{3}{16\pi} \dfrac{m_c^2}{v^2}
\approx \pm 3 \cdot 10^{-7} ~,
$$
where we assumed $|M_3|/|M_{1,2}|\gtrsim 1.5$, 
so that the effects of the three-neutrino degeneracy can be disregarded. 
Let us estimate
the contribution to $\epsilon_{1,2}$ of diagrams with $N_{2,1}$ in the loop.  
The maximal asymmetry is obtained when 
\beq
\dfrac{|M_2|-|M_1|}{|M_1|}\approx\dfrac{\Gamma_{1,2}}{|M_1|}\approx 
\dfrac{m_t^2}{16\pi v^2} ~.
\label{res}
\end{equation}
In this case the function $f$ in Eq. (\ref{epsi}) should be replaced by
$|M_1|/(2\Gamma_1)$ \cite{pila}, and one finds
\beq
\epsilon_1\approx\epsilon_2\approx
\dfrac{1}{16\pi}\dfrac{|M_1|}{\Gamma_1}
\dfrac{{\rm Im}(h^\dag h)^2_{12}}{(h^\dag h)_{11}}
\approx \dfrac 12 \sin(\phi'_1 - \phi'_2 ) ~.
\label{eps3}
\end{equation}
As in the special case I, the factor $ \sin(\phi'_1 - \phi'_2)$ is suppressed 
because of the approximately opposite CP parities of $N_1$ and $N_2$
(see Eq. (\ref{phases})). Computing also ${\cal O}(\lambda^2)$ terms in 
$U_R$, we find  
\beq
\sin(\phi'_1 - \phi'_2 )\sim\dfrac{m_u}{m_t}\approx 10^{-5}~.
\label{sin}
\end{equation}
As far as $\epsilon_3$ is concerned, the two contributions proportional to
Im$(h^\dag h)_{31}^2$ and Im$(h^\dag h)_{32}^2$ are of order
$m_t^2/(16\pi v^2)$, but have opposite sign because of the opposite
CP parities of $N_1$ and $N_2$. Moreover, $\epsilon_3$ is washed out
efficiently by the strong L-violating interactions of $N_1$ and $N_2$. 
Therefore its contribution to $\eta_B$ can be neglected and
we finally obtain
$$
\eta_B \approx 0.01 \cdot 2 \epsilon_1\kappa_1
\sim 10^{-14}\left(\dfrac{10^5m_u}{m_t}\right)
\left(\dfrac{\kappa_1(m_u/m_t)}{10^{-7}}\right)
~.
$$
Thus, the leptogenesis is not successful in this special case.

Reducing the ratio $m_t/m_u$, one gets both smaller washout and enhanced
asymmetries $\epsilon_{1,2}$ (see Eqs. (\ref{ti3}), (\ref{eps3}) and 
(\ref{sin})). However, to obtain $\eta_B$ in the correct range, one would 
have to violate the assumption in Eq. (\ref{HD}).  
Even a strong degeneracy between all three RH neutrinos cannot lead to a 
sufficient increase of the final baryon asymmetry, because the 
enhancement due to the degeneracy is limited by the large values of 
$\Gamma_{1,2}$ given in Eq. (\ref{res}) (see also 
the discussion at the end of this section). 
Moreover, the analytic approximation (\ref{k1}) most probably
underestimates the washout effects in this case, because of the very large
values of $\tilde{m}_1$ and $\tilde{m}_2$ ($\sim 500$ eV).
To the best of our knowledge, no numerical solutions of the relevant
Boltzmann equations in this regime are available in the literature, since
it is usually assumed that $\tilde{m}_1$ does not exceed the mass of the
heaviest left-handed neutrino $|m_3|$. The present special case shows
that this is not always true.

(b) $r>1$, heavy quasi-degenerate pair.

In this case $\epsilon_1$ gives the dominant contribution to the final 
baryon asymmetry. 
The RH mixing matrix is 
obtained from that in Eq. (\ref{ura}) by the cyclic permutation of its
columns $3\to 1$, $1\to 2$, $2\to 3$. 
Using the approximation
$$
\delta_2\approx\dfrac{m_c}{m_t}
\dfrac{r^2}{r^2-1} e^{-i\phi/2}~,
$$
we obtain from Eq. (\ref{hh}) 
$$
\begin{array}{l}
(h^\dag h)_{11}\approx\dfrac{m_c^2}{v^2}\dfrac{1-2r^2+2r^4}{1-2r^2+r^4}~,\\
(h^\dag h)_{12}\approx -e^{i(\phi'_3-\phi'_2)/2}(h^\dag h)_{13} 
\approx \dfrac{m_t m_c}{\sqrt{2}v^2}\dfrac{r^2}{1-r^2}
e^{i(\phi+\phi'_1-\phi'_2)/2}
~.
\end{array}
$$
Then for $\tilde{m}_1$ and $\kappa_1$ we find
$$
\tilde{m}_1 \gtrsim 2|m_{\mu\mu}|\approx 0.05 ~{\rm eV}~,~~
~~~~\kappa_1(\tilde{m}_1)\lesssim 3\cdot 10^{-3}~.
$$
The asymmetry produced in the decays of $N_1$ can be written as
$$
\epsilon_1 \approx -\dfrac{3m_t^2}{32\pi v^2}
\dfrac{r^3}{1-2r^2+2r^4}\sin\psi\sin(\phi'_3-\phi'_2)
\approx 3 \cdot 10^{-3} ~\sin(\phi'_3-\phi'_2)~,
$$
where $\psi\equiv(\phi+\phi'_1-\phi'_2/2-\phi'_3/2)$.
In the last equality we have chosen $r=2$ and $\sin\psi =1$, which 
are the most favorable values for obtaining a large $\eta_B$ (note that, 
even though $\epsilon_1$ is maximized at $r\simeq 1$, in the limit 
$r\rightarrow 1$ the parameter $\tilde{m}_1$ becomes very large, which
signals a very strong washout of the asymmetry).
Also in this case the fact that the CP parities of $N_2$ and $N_3$ 
are almost opposite leads to a strong cancellation:
$$
\sin(\phi'_3-\phi'_2)\sim\dfrac{m_u}{m_t}\approx 10^{-5}~.
$$
Thus we obtain
$$
\eta_B \approx 0.01 \cdot  \epsilon_1 \cdot \kappa_1 
\lesssim 5\cdot 10^{-13}\left(\dfrac{m_u}{1{\rm MeV}}\right)
\left(\dfrac{m_t}{100{\rm GeV}}\right)
~.
$$
Increasing $m_u \cdot m_t$ (and also $m_c^2$ in 
order to keep $r$ fixed) would increase $\eta_B$. 

However, it is unlikely that this would lead to a successful leptogenesis.  
Indeed, since all three RH neutrino masses are of the same order in this 
special case ($r\sim 1$), the heavier neutrinos $N_2$ and $N_3$ are still 
abundant at the temperature $T\sim |M_1|$ at which the decays of $N_1$ take 
place. Therefore the strong washout effects due to the processes involving 
$N_2$ and $N_3$, which are characterized by very large $\tilde{m}_{2,3}\simeq
500$ eV, are expected to efficiently wash out the asymmetry $\epsilon_1$ 
even though the parameter $\tilde{m}_1$ is relatively small. 
Therefore we do not 
expect this special case to lead to a successful baryogenesis through 
leptogenesis. A more accurate study of the case of three quasi-degenerate 
RH neutrinos would require solving numerically a coupled set of Boltzmann 
equations describing the evolution of the number densities of all RH
neutrinos and $B-L$. We consider such a study, which is beyond the scope
of the present paper, to be very desirable.

\section{Discussion and conclusions \label{con}}

We have analyzed the possibility of explaining both the low energy neutrino 
data and the observed baryon asymmetry of the Universe in the framework of 
the seesaw mechanism and studied the requisite structure of the RH neutrino 
sector. Our analysis was based on the assumptions of hierarchical 
eigenvalues of the Dirac mass matrix $m_D$ and small Dirac-type 
left-handed mixing ($U_L\approx\mathbb{1}$).

Let us now abandon the hypothesis $U_L\approx\mathbb{1}$. If the matrix 
$U_L$ is arbitrary, the direct connection between the low energy data
and the structure of $M_R$ is lost. 
This additional freedom relaxes the phenomenological constraints on 
RH neutrinos. In fact, now the unique low energy requirement on the seesaw 
mechanism is to reproduce the light neutrino masses, given by the eigenvalues 
of $\hat{m}$ (see Eq. (\ref{tilm})); the correct leptonic mixing matrix 
$U_{PMNS}$ can always be obtained through the proper choice of $U_L$. 
As an example, let us consider the case of non-degenerate RH masses and take 
the following RH mixing matrix:
$$
U_R=\left(
\begin{array}{ccc}
\dfrac{1}{\sqrt{2}} & \dfrac{1}{\sqrt{2}} & 0 \\
-\dfrac{1}{\sqrt{2}} & \dfrac{1}{\sqrt{2}} & 0 \\
0 & 0 & 1
\end{array}\right)\cdot K ~.
$$
The maximal mixing of the two lighter RH neutrinos will maximize the lepton
asymmetry. The eigenvalues of the matrix
$$
\hat{m}=-m_D^{diag} W m_D^{diag}
$$
(see Eq. (\ref{mr-1})) are given, approximately, by $m_c^2/(4|M_2|),~m_c^2/
(2|M_1|),~m_t^2/|M_3|$. Taking $|M_1|\approx  10^{10}$ GeV$\cdot
(m_c/0.4$ GeV$)^2$, $|M_2|$ a few times larger and $|M_3|\approx 2\cdot 
10^{14}$ GeV $\cdot(m_t/100~{\rm GeV})^2$, one can reproduce the solar 
and atmospheric mass squared differences. Since $\hat{m}$ is approximately 
diagonal, the solar and atmospheric mixing angles are generated by $U_L$, which 
should have an almost bimaximal form. 

It is easy to calculate the washout mass parameter and the asymmetry produced 
in the decays of $N_1$:
$$
\tilde{m}_1=\dfrac{m_c^2}{2|M_1|}\approx \sqrt{\Delta m^2_{sol}}~,~~~~~
\epsilon_1 \approx \dfrac{3m_c^2}{32\pi v^2}\sin(\phi_2-\phi_1)
\dfrac{|M_1|}{|M_2|}~.
$$
Assuming $\phi_2-\phi_1\sim \pi/2$ (note that the CP parities of $N_1$ and 
$N_2$ are in general not constrained), we get
$$
\eta_B \approx 3\cdot 10^{-11}\dfrac{|M_1|}{|M_2|}\left(\dfrac{m_c}
{0.4~{\rm GeV}}\right)^2 ~.
$$
Thus, for a moderate hierarchy between $M_1$ and $M_2$, a value of $m_c$ 
around a few GeV can lead to a successful leptogenesis. 
This example shows that, relaxing the hypothesis $U_L\approx \mathbb{1}$,
it is easier to realize baryogenesis via leptogenesis. In particular, 
the degeneracy of the masses of RH neutrinos $|M_i|$ is no longer necessary, 
but the hierarchy of $|M_i|$ should not be as large as it is in the generic 
case. 

Let us now discuss the renormalization group equation (RGE) evolution of the 
neutrino mass matrices. 
The structure of the effective mass matrix $m$ is stable under the Standard 
Model (or MSSM) radiative corrections \cite{CP}. The corrections to its 
matrix elements can be written as
$$
\Delta m_{\alpha\beta} \approx (\epsilon_\alpha + \epsilon_\beta) 
m_{\alpha\beta} ~,
$$
where $\epsilon_\alpha$ ($\lesssim 10^{-2}$) describes the effect of the Yukawa 
coupling of the charged lepton $l_\alpha$. Therefore both $m_{ee}$ and 
$d_{12}\equiv (m_{ee}m_{\mu\mu}-m_{e\mu}^2)$ receive small corrections 
proportional to themselves: if $m_{ee}$ and/or $d_{12}$ are very small at 
the electroweak scale, they remain very small also at the seesaw scale 
(the mass scale of RH neutrinos), and so the level crossing conditions do not 
change.

Between the GUT and the seesaw scales one has to consider the evolution of 
the neutrino Yukawa couplings and of Majorana 
mass matrix of RH neutrinos rather than the evolution of the effective 
matrix $m$ \cite{HOS}. We assume that at the GUT scale the Yukawa couplings of 
neutrinos $h$ are related with those of quarks or charged leptons. 
The evolution of $h$ with decreasing mass scale will not modify the 
hierarchy $m_u\ll m_c \ll m_t$, and its effects can be absorbed into a 
redefinition of our indicative values of $m_{u,c,t}$.

The RGE effects on $M_R$ are due to the neutrino Yukawa couplings; they 
can, in principle, be important in the cases of strongly degenerate RH 
neutrinos. Consider the stability of the structure of $M_R$ in the special 
case that leads to a successful leptogenesis. Recall that in this case the 
$(12)$-sector of RH neutrinos is characterized by $|M_{1,2}|\approx 10^8$ 
GeV, $(|M_2|-|M_1|)/|M_1|\lesssim 10^{-5}$ and $\Delta\approx \pi/2$, where 
$\Delta$ is defined in Eq. (\ref{k}). The largest correction to the 
$(12)$-block of $M_R$ between $M_{\rm GUT}$ and $|M_{1,2}|$ is the 
correction to the $22$-element: 
$$
\dfrac{(\Delta M_R)_{22}}{(M_R)_{22}}\sim
\dfrac{m_c^2}{16\pi^2 v^2}
\log\left(\dfrac{M_{\rm GUT}}{10^8{\rm~GeV}}\right) \approx 6\cdot 10^{-7}
\left( \dfrac{m_c}{0.4{\rm~GeV}}\right)^2 ~.
$$
Therefore, the radiative corrections cannot generate a relative splitting 
between $|M_1|$ and $|M_2|$ exceeding  $10^{-5}$. Moreover, at one loop level, 
the phases of $(M_R)_{ij}$ have no RGE evolution and so the relation 
$\Delta\approx\pi/2$ is not modified.


Let us summarize the main results of our analysis:

1) We have discussed the properties of the seesaw mechanism under the 
assumption of an approximate quark-lepton symmetry, which implies a similarity 
between the Dirac neutrino mass matrix $m_D$ and the mass matrices of charged 
leptons and quarks. This, in turn, implies a strong hierarchy of the eigenvalues 
of $m_D$ and small left-handed Dirac-type mixing.

2) The presence of two large mixing angles ($\theta_{12}$ and $\theta_{23}$) 
and relatively weak mass hierarchy of light neutrinos lead, in general, to 
a ``quasi-democratic" structure of the mass matrix $m$ in the flavor basis, 
with values of all its elements within one order of magnitude of each other. 
A strong hierarchy of the elements appears in special cases only.  

3) In the generic case (nearly democratic $m$), the mass matrix of RH
neutrinos has a strong (nearly quadratic in $m_D$) hierarchy of eigenvalues 
and small mixing. The lightest RH neutrino has a mass $|M_1| < 10^6$ GeV, 
well below the absolute lower bound coming from the condition of a successful 
leptogenesis. As a result, the predicted lepton asymmetry is smaller than 
$10^{-14}$, and the scenario of baryogenesis via leptogenesis  does not 
work. 
 
4) The special cases correspond to the level crossing points, when either 
two or all three masses of RH neutrinos are nearly equal. We have found two 
level crossing conditions: (1) $m_{ee}\rightarrow 0$ (the $N_1 - N_2$ crossing)
and  (2) $d_{12} \rightarrow 0$ ($N_2 - N_3$ crossing), where $m_{ee}$
and $d_{12}$ should be evaluated in the basis where the Yukawa couplings 
of neutrinos are diagonal. In the crossing points the mixing of the 
corresponding neutrino states is maximal and their CP parities are nearly 
opposite.

5) For $U_L\simeq \mathbb{1}$ the leptogenesis can be successful only 
in the special case with small element $m_{ee}$, which corresponds to the  
$N_1 - N_2$ crossing. It is characterized by $|M_1| \approx |M_2| \approx 
10^{8}$ GeV, $|M_3| \approx  10^{14}$ GeV and $(|M_2| - |M_1|)/|M_2| \lesssim  
10^{-5}$. $N_1$ and  $N_2$ are strongly mixed and their mixing with $N_3$ is 
very small. The CP-violating phase $\Delta$ in Eq. (\ref{k}) should be very 
close to $\pi/2$. 
Notice that this unique 
case with a successful leptogenesis is defined very precisely. It has a 
number of characteristic features which can give important hints for model 
building. 

6) For $U_L=\mathbb{1}$, the successful scenario is realized for the normal 
mass hierarchy of light neutrinos and predicts a very small effective Majorana 
mass probed in the neutrinoless $2\beta$-decay: $|m_{ee}| \lesssim 10^{-4}$ 
eV. However, for $U_L\approx U_{CKM}$, this case can be realized also for 
other mass spectra and  $|m_{ee}|$ as large as $\sim 0.1$ eV.

7) We find that low-energy neutrino data allow also the other special cases, 
with $2-3$ crossing or both $1-2$ and $2-3$ crossings of the masses of RH 
neutrinos. These cases, however, do not lead to a successful baryogenesis 
through leptogenesis.

The seesaw mechanism can account for both the low-energy neutrino data
and a successful thermal leptogenesis, but a very specific structure of
the mass matrices is required. Although this structure may look as an 
extreme fine tuning when viewed from the low-energy (effective theory) 
side ($m_{ee}\to 0$, $|m_{\mu\mu}|\approx (m_u/m_c)^2|m_{ee}|$),
it does not appear unnatural from the point of view of the fundamental 
physics responsible for the seesaw mechanism: indeed, it just requires 
an approximate degeneracy and nearly maximal mixing of the two lightest 
RH neutrinos, which may well be a consequence of some flavor symmetry 
operating in the RH sector.

Can the unique successful special case that we found be ruled out? Since it 
requires a suppression of $|m_{ee}|$, it will be excluded in case of a 
positive signal of $2\beta 0\nu$-decay with $|m_{ee}|$ close to the 
heaviest of the light neutrino masses (which could be measured in direct 
neutrino mass search experiments). In that case one will be left with the 
following alternatives:

\begin{itemize} 

\item the quark-lepton symmetry is strongly violated: there is
no strong hierarchy of the eigenvalues of $m_D$ 
($m_u/m_c,m_c/m_t \gtrsim 10^{-1}$) and/or the Dirac-type left-handed mixing
is large (the corresponding mixing angles 
are larger than $\theta_{c} \approx 0.2$);
\item type-I seesaw \cite{yana}
is not the sole source of neutrino mass; the simplest alternative could be 
type-II seesaw \cite{mose2} in which there is an additional contribution 
from an $SU(2)_L$-triplet Higgs. Another possibility is that 
the seesaw is not the true mechanism of neutrino mass generation;

\item a mechanism other than the decay of thermally produced
RH neutrinos contributes to 
leptogenesis or the baryon asymmetry of the Universe is generated 
through a different mechanism, which has nothing to do with leptogenesis.\\

\end{itemize}

\noindent{\bf Note added}.

Let us comment on the possibility of non-thermal production of
the heavy RH neutrinos (see, e.g., \cite{muya}), 
that in principle can lead to a successful leptogenesis for values of
the parameters $M_1$ and $\tilde{m}_1$ for which thermal leptogenesis 
does not work.

In fact, it is interesting that also non-thermal leptogenesis is strongly 
constrained in our framework. Consider the generic case (section 3).
Since $M_1$ is relatively light ($\lesssim 10^7$ GeV), 
$\epsilon_1$ is very small.
Moreover, as $\tilde{m}_1$ is relatively large 
($\gtrsim \sqrt{\Delta m^2_{sol}}$),
the washout effects suppress (at least partially) the asymmetry
generated in the decays of non-thermally produced RH neutrinos
\cite{snEll}.
As a consequence, even in the non-thermal case, 
the asymmetry generated by $N_1$ turns out to be insufficient and, 
to enhance it, one has
to resort again to the special case I (section 5). 

It is known, however (see, e.g., \cite{ANT}), that also the
asymmetries generated by $N_2$ and/or $N_3$ can survive if 1) 
they are produced non-thermally at reheating and 2)
$N_1$ is not in thermal equilibrium at the reheating
temperature $T_{RH}$.
In fact, the asymmetries $\epsilon_{2,3}$ can be large (they are of the order 
of $m_{c,t}^2/(16\pi v^2)$ in the generic case and even larger in the
special case II: $\epsilon_{2,3}\sim m_c/m_t$).
However, partial thermalization of $N_{2,3}$ and subsequent washout 
can occur after reheating. Moreover, $N_1$ should not enter into thermal
equilibrium at any temperature $T\lesssim T_{RH}$.

In this case an accurate computation of the final asymmetry 
would require to solve the complete set
Boltzmann equations describing the evolution of the
number densities of all three RH neutrinos and of $B-L$.

\section*{Acknowledgments}

We are grateful to L. Boubekeur, W. Buchm\"uller, W. Rodejohann and 
Y. Takanishi for useful discussions. E.A. acknowledges 
the warm hospitality of the Kavli Institute for Theoretical Physics, 
University 
of California, Santa Barbara, where this paper was completed.  The research 
of E.A. was supported in part by the National Science Foundation under Grant
No. PHY99-07949.  
The work of M.F. is supported in part by the Italian MIUR under the program
``Fenomenologia delle Interazioni Fondamentali'' and by INFN under the program
``Fisica Astroparticellare''.

\bibliographystyle{JHEP}
\bibliography{biblio}

\end{document}